\documentclass[iop]{emulateapj}
\pdfoutput=1

\usepackage{bm}
\usepackage{array}
\usepackage{amsmath}
\usepackage[usenames,dvipsnames]{xcolor}
\usepackage{overpic}

\usepackage[backref,colorlinks,citecolor=blue]{hyperref}        
\usepackage[all]{hypcap}

\defcitealias{Bonsor15}{Paper I}

\newcommand{\icarus}{Icarus}

\shorttitle{Compositional evolution of rocky protoplanets}
\shortauthors{Carter et al.}

\begin{document}

\title{Compositional evolution during rocky protoplanet accretion}

\author{Philip J. Carter, Zo\"{e} M. Leinhardt}
\affil{School of Physics, University of Bristol, H. H. Wills Physics Laboratory, Tyndall Avenue, Bristol BS8 1TL, UK}
\email{p.carter@bristol.ac.uk}

\author{Tim Elliott, Michael J. Walter}
\affil{School of Earth Sciences, University of Bristol, Wills Memorial Building, Queen's Road, Bristol BS8 1RJ, UK}

\and

\author{Sarah T. Stewart}
\affil{Department of Earth and Planetary Sciences, University of California, One Shields Avenue, Davis, CA 95616, USA}

%%%%%%%%%%%%%%%%%%%% 			ABSTRACT 		%%%%%%%%%%%%%%%%%%%%

\begin{abstract}
The Earth appears non-chondritic in its abundances of refractory lithophile elements, posing a significant problem for our understanding of its formation and evolution. It has been suggested that this non-chondritic composition may be explained by collisional erosion of differentiated planetesimals of originally chondritic composition. In this work, we present $N$-body simulations of terrestrial planet formation that track the growth of planetary embryos from planetesimals. We simulate evolution through the runaway and oligarchic growth phases under the Grand Tack model and in the absence of giant planets. These simulations include a state-of-the-art collision model which allows multiple collision outcomes, such as accretion, erosion, and bouncing events, that enables tracking of the evolving core mass fraction of accreting planetesimals. We show that the embryos grown during this intermediate stage of planet formation exhibit a range of core mass fractions, and that with significant dynamical excitation, enough mantle can be stripped from growing embryos to account for the Earth's non-chondritic Fe/Mg ratio. We also find that there is a large diversity in the composition of remnant planetesimals, with both iron-rich and silicate-rich fragments produced via collisions.
\end{abstract}


%%%%%%%%%%%%%%%%%%%% 		INTRODUCTION 		%%%%%%%%%%%%%%%%%%%%

\section{Introduction}

{ As the best studied planetary system, our own solar system provides the most important benchmark for theories of planet formation.  Critically, we have detailed compositional information on the Solar system from spectroscopic measurements of the sun and analyses of meteorites \citep[e.g.][]{Lodders03}.  By assuming that planets have the composition of primitive meteorites, at least for refractory, lithophile elements that vary little between meteorite groups,  inferences about planetary evolution can be made from the deviation of observed compositions from such a likely starting point.  In the last decade, the validity of the notion of a chondritic bulk planetary composition, has faced close scrutiny \citep[e.g.][]{BoyetandCarlson05} and several studies have suggested that the process of accretion may inevitably generate planets with non-chondritic compositions \citep[e.g.][]{Bourdon08,ONeill08}. However, these ideas have yet to be explored in detail with physical models.

The terrestrial planets of our solar system exhibit a range of Fe/Mg ratios. The Earth (Fe/Mg$\,\simeq\,$2.1, \citealt{ONeill08}) is significantly enriched in iron compared to both carbonaceous chondrites (1.92$\pm$0.08, \citealt{Palme+Jones03}) and the solar photosphere (1.87$\pm$0.4, \citealt{Lodders03}), and Mercury with its massive core must have a large iron excess (estimated Fe/Mg\,=\,9.92, \citealt{Morgan80}). Estimates for Venus \citep{Morgan80} and Mars \citep{Lodders97} suggest they have Fe/Mg ratios similar to the Earth and chondrites respectively. In contrast the Moon is expected to have a Fe/Mg ratio significantly lower than chondritic values, and Fe/Mg ratios for meteorites (including differentiated bodies) range from less than 0.08 to greater than 25 \citep{Nittler04}. Such extremes would not be expected for primitive bodies; iron meteorites, for example, probably represent the surviving cores of differentiated planetesimals. 
It has been proposed that stripping of crust and mantle via collisions during accretion led to this observed compositional variation \citep[e.g.][]{Benz88,ONeill08}.

Recently some progress has been made by use of $N$-body accretion models which allow for imperfect merging \citep{Bonsor15,Dwyer15}.  Both of these studies investigated discs unperturbed by giant planets, the calmest possible scenario for the formation of terrestrial planets.  Migration of the giant planets has been argued to play a key role in explaining the small size of Mars, depopulating the asteroid belt and scattering potentially volatile rich material into the inner Solar system \citep[e.g.][]{Walsh11}.  The excitation of planetesimals such migration causes may also have significant implications for the importance of collisional erosion during the accretion of the terrestrial planets.

In this work, we extend the work of \citet{Bonsor15} (hereafter \citetalias{Bonsor15}) in assessing the likelihood of non-chondritic accretion of terrestrial planets, by examining the dynamically hot scenario provided by the Grand Tack model \citep{Walsh11}.  We compare this to an updated model of calm disc accretion, which additionally explores the effects of gas drag.}

%%%%%%%%%%%%%%%%%%%% 			METHODS 		%%%%%%%%%%%%%%%%%%%%

\section{Numerical Method}

We use a modified version of the parallelized $N$-body code PKDGRAV \citep{Richardson00,Stadel01} to simulate the gravitational interactions of planetesimals in the potential well of a star. A state-of-the-art collision model \citep[EDACM,][]{Leinhardt12,Leinhardt15} has been incorporated into this code in order to calculate the results of various collision types, including partial accretion, hit and run and erosive events. { PKDGRAV allows us to simulate much larger numbers of particles than codes used in similar works \citep[e.g.][]{Walsh11,Chambers13,OBrien14}, and to include planetesimal-planetesimal interactions. However, the second-order integrator would not achieve sufficient accuracy for the long term orbital evolution of embryos during the giant impact phase -- the final stage of planet formation characterised by stochastic collisions between large embryos that have developed chaotic orbits due to their mutual interaction. In this work we simulate the intermediate stages of  planet formation characterised by runaway and oligarchic growth of planetesimals and embryos.} All simulations were carried out using the University of Bristol's BlueCrystal\footnote{http://www.bris.ac.uk/acrc/} supercomputer.

\subsection{Initial planetesimal disc}

In this work we present simulations initiated with either $10\,000$ (low resolution) or $100\,000$ (high resolution) particles arranged in an annulus, extending from 0.5\,AU to either 1.5 or 3\,AU, around a one solar mass star.

The initial disc of planetesimals is distributed according to a surface density, $\Sigma = \Sigma_1 a^{-3/2}$, similar to that of the Minimum Mass Solar Nebula (MMSN, \citealt{Weidenschilling77,Hayashi81}), using a surface density at 1\,AU of $\Sigma_1 = 10$\,g\,cm$^{-2}$ to match previous work (e.g.\ \citealt{Kokubo+Ida02,Leinhardt09}; \citetalias{Bonsor15}).

\begin{figure*}
\centering
\begin{overpic}[width=\textwidth]{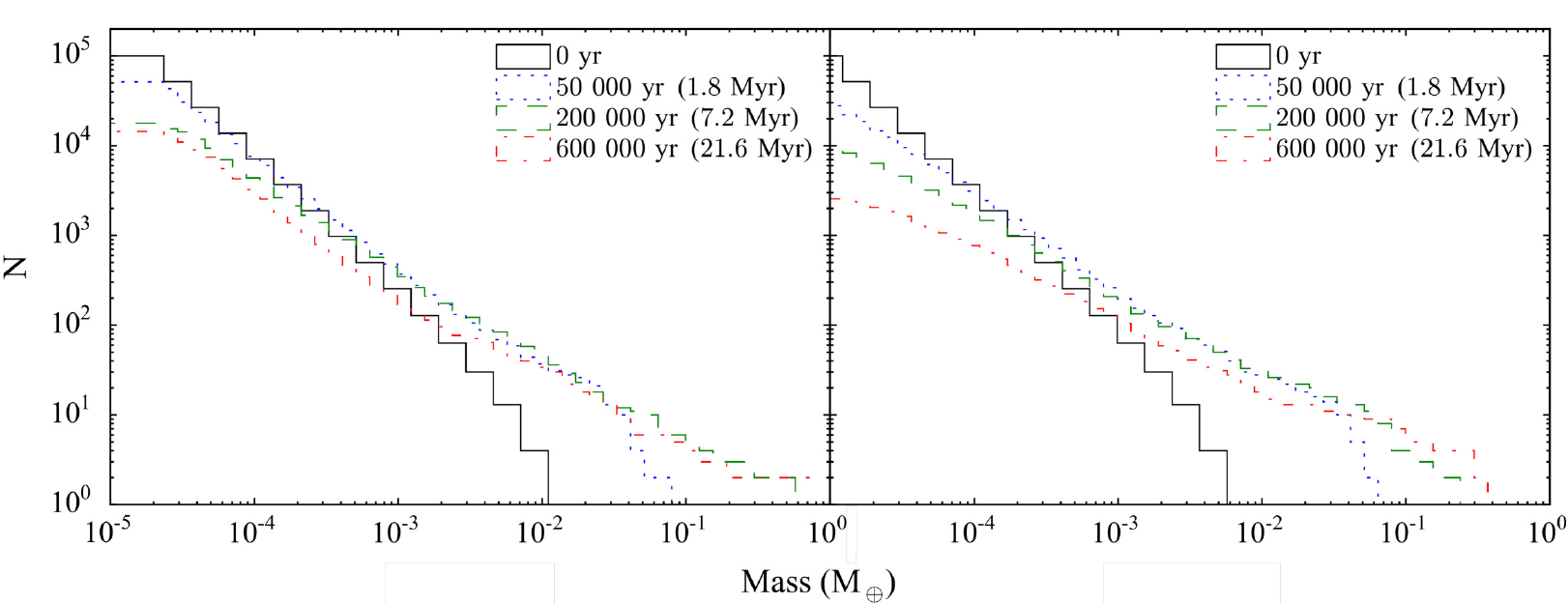}
\put(8,8){\normalsize GT}
\put(54,8){\normalsize Calm}
\end{overpic}
\caption{Evolution of the planetesimal mass distribution (cumulative number of bodies) for a high resolution simulation under the Grand Tack scenario (simulation \ref{022GTJf6nogas}, left) and a calm disc simulation (\ref{022f6nogas}, right). \label{f:massdist}}
\end{figure*}

We begin our simulations with planetesimals of 15 different sizes, evenly spaced in mass, $m$, and chosen according to the cumulative number distribution:
\begin{equation}
\mathrm{d}N = m^{-p} \mathrm{d}m,
\end{equation}
with $p = 2.5$, a slope appropriate for the runaway growth phase \citep{Kokubo+Ida96}. This initial mass distribution (black line in Figure \ref{f:massdist}) is similar to that used in \citetalias{Bonsor15} (in which the size distribution was generated by running PKDGRAV using perfect merging with a disc with five times as many equal size planetesimals). The simulations begin with most of the planetesimals having radii of a few hundred km { in accordance with the mass distribution shown in Figure \ref{f:massdist}}.

In this work we study the intermediate stage of planet formation, sizeable planetesimals have already formed and will continue to grow into embryos via runaway accretion during the simulations. As it is known that planetesimals can differentiate in the first few million years of the Solar System's evolution { \citep[e.g.][]{Lugmair98,Srinivasan99,Amelin08,Kruijer14}} we assume that all planetesimals are differentiated at the start of our simulation, and assign the same core fraction to all particles, regardless of their location within the disc. This uniform initial differentiation is probably a simplification, but will aid understanding of the role of collisions on the compositional evolution of growing bodies.

The initial conditions used for the simulations presented in this paper are summarised in Tables \ref{t:initconds} and \ref{t:initcondsGT}.
\begin{table*}
\centering
\caption{Initial conditions for the calm disc simulations.\label{t:initconds}}
\begin{tabular}{l r r}
\hline
Parameter &			 			Value & 													Notes \\
\hline
Mass of planetesimal disc	 &			2.5\,M$_\oplus$ &											\\
Radial extent of disc &				0.5-1.5\,AU &												\\
Initial number of particles & 			100\,000 & 												High resolution \\
 &								10\,000 &													Low resolution \\
Initial planetesimal mass &			$3.2 \times 10^{-11}$\,M$_\odot$ - $1.5 \times 10^{-8}$\,M$_\odot$ & 	High resolution \\
 &								$3.2 \times 10^{-10}$\,M$_\odot$ - $3.6 \times 10^{-8}$\,M$_\odot$ &	Low resolution \\
Initial planetesimal radius &			196\,km - 1530\,km &				 						High resolution ($f=1$) \\
 &								423\,km - 2050\,km &										Low resolution ($f=1$) \\
Initial core fraction & 				0.22, 0.35 & 												\\
Planetesimal density &				2\,g\,cm$^{-3}$ &											\\
Resolution limit &					$3 \times 10^{-11}$\,M$_\odot$ ($1 \times 10^{-5}$\,M$_\oplus$) &	 	High resolution \\
 &								$3 \times 10^{-10}$\,M$_\odot$ ($1 \times 10^{-4}$\,M$_\oplus$) &	 	Low resolution \\
Timestep &						0.01\,yr, 0.005\,yr (with gas drag) &								\\
Duration &						600\,000\,yr ($\sim$20\,Myr effective time) &						\\
Initial gas density, $\rho_\mathrm{g,1}$ &	$1.4 \times 10^{-9}$ g\,cm$^{-3}$ &								MMSN \\
Gas dissipation start time &			56\,kyr (2\,Myr effective time) & 									 \\
Gas dissipation timescale &			2.8\,kyr (100\,kyr effective time), $\infty$ & 						 \\
\hline
\end{tabular}
\end{table*}
\begin{table*}
\centering
\caption{Initial conditions for the Grand Tack simulations.\label{t:initcondsGT}}
\begin{tabular}{l r r}
\hline
Parameter &			 			Value & 													Notes \\
\hline
Mass of planetesimal disc	 &			4.85\,M$_\oplus$ &											\\
Radial extent of disc &				0.5-3.0\,AU &												\\
Initial number of particles & 			100\,000 & 												High resolution \\
 &								10\,000 &													Low resolution \\
Initial planetesimal mass &			$6.2 \times 10^{-11}$\,M$_\odot$ - $2.9 \times 10^{-8}$\,M$_\odot$ & 	High resolution \\
 &								$6.2 \times 10^{-10}$\,M$_\odot$ - $7 \times 10^{-8}$\,M$_\odot$ &	Low resolution \\
Initial planetesimal radius &			245\,km - 1900\,km &				 						High resolution ($f=1$) \\
 &								528\,km - 2550\,km &										Low resolution ($f=1$) \\
Initial core fraction & 				0.22, 0.35 & 												\\
Planetesimal density &				2\,g\,cm$^{-3}$ &											\\
Resolution limit &					$6 \times 10^{-11}$\,M$_\odot$ ($2 \times 10^{-5}$\,M$_\oplus$) &	 	High resolution \\
 &								$6 \times 10^{-10}$\,M$_\odot$ ($2 \times 10^{-4}$\,M$_\oplus$) &	 	Low resolution \\
Timestep &						0.01\,yr, 0.005\,yr (with gas drag) &								\\
Duration &						600\,000\,yr ($\sim$20\,Myr effective time) &						\\
Jupiter initial location &				3.5\,AU &													\\
Jupiter minimum semi-major axis &		1.5\,AU &													\\
Jupiter final location &				5.2\,AU &													\\
Migration timescale &				2.8\,kyr (100\,kyr effective time), 17\,kyr (600\,kyr) & 					Fast, slow migration \\
Migration start time &				56\,kyr (2\,Myr effective time), 220\,kyr (8\,Myr) & 					Early, late migration \\
Initial gas density, $\rho_\mathrm{g}$ &	$1.4 \times 10^{-9}$ g\,cm$^{-3}$ &								MMSN \\
 &								see equations \eqref{hydrogas} \& \eqref{hydrogas2} &				Based on hydrodynamic simulations \\
\hline
\end{tabular}
\end{table*}

\subsection{Expansion factor}

$N$-body simulations of the intermediate phase of planet formation with large $N$ have often employed an expansion of the particle radii in order to increase the collision rate, and hence speed up the evolution, e.g.\ \citet{Kokubo+Ida96,Kokubo+Ida02,Leinhardt05,Morishima08}. This expansion factor was discussed in detail by \citet{Kokubo+Ida96,Kokubo+Ida02}, in which they determine that it has very little effect on the evolution during this phase, merely altering the timescale for planet formation. { In this work we are studying the combined consequences of many collisions of different types, including erosive collisions, and since the expansion factor prevents very close approaches from occurring (which would excite planetesimal velocities), it is expected that our simulations slightly underestimate mantle erosion.

While the number of bodies in the simulation is large, dynamical friction from the planetesimals controls the velocity dispersion, and the expansion factor does not alter their growth mode \citep{Kokubo+Ida96}.  As the number of planetesimals decreases, and massive embryos begin to significantly alter velocities via gravitational scattering, the expansion factor likely causes the results to become less accurate. Once the giant impact stage of planet formation begins this becomes a significant problem, and so we cannot integrate accurately through the late stages of formation using the expansion factor.} Due to limitations in computing power it is still not practical to conduct large $N$ simulations covering millions of years of planet formation without using this expansion factor. 

In previous work we have used an expansion factor, $f = 6$ (\citealt{Leinhardt05,Leinhardt09}; \citetalias{Bonsor15}; \citealt{Leinhardt15}), and, for consistency, we adopt this same value here. In our simulations giant planets do not have their radii inflated (any planetesimals colliding with giant planets are forced to merge to avoid complications), and gas drag is calculated using the natural radius, not the expanded value.

During the intermediate stages of planet formation we are investigating, the expansion factor causes the evolution to speed up by a factor $\sim f^2$. We therefore quote both a simulation time, and an \emph{effective} time corresponding to realistic radii ($f=1$).

\subsection{Collision model and debris}

EDACM \citep{Leinhardt12,Leinhardt15} provides an analytical description of the outcome of a collision between gravity-dominated bodies, like the planetesimals modelled in this work. Using this collision model the size and velocity distributions of the fragments of each collision are calculated based on the impact parameter and collision velocity, with each collision classified as having one of seven different outcomes: perfect merging, partial accretion, hit and run (with the projectile intact, disrupted, or supercatastrophically disrupted), erosion, or supercatastrophic disruption.

Since EDACM can produce many fragments from the original two impactors it is necessary to impose a resolution limit, a mass below which no further resolved particles will be generated. Any remnants smaller than this limit are treated semi-analytically as `unresolved debris'. This debris is distributed in a series of circular annuli each 0.1\,AU wide, with the debris placed in the annulus corresponding to the location of the collision that produced it. This material is assumed to have circular Keplerian orbits. As well as being generated by collisions, this unresolved debris is reaccreted by the resolved particles according to their geometric cross-section (calculated using inflated radii) and orbit eccentricity (see \citealt{Leinhardt05}). As was shown by \citet{Leinhardt15}, there is never a significant amount of mass in this unresolved component (see also section \ref{s:debris}).

In this work we track both the total mass of debris in each annulus, and the mass of core material (since this proved to be a problem in \citetalias{Bonsor15}). Additionally, each particle and debris annulus has an `origin histogram', that tracks the fraction of that object's mass originating from within each 0.1\,AU annulus (see \citealt{Leinhardt05} for details).

\subsection{Mantle stripping law}\label{s:initialcf}

Differentiated planetesimals are modelled with core and mantle components, which are tracked independently. In a collision between two such planetesimals core { and} mantle material may be exchanged, or distributed amongst fragments. \citet{Marcus10} analysed SPH collisions between differentiated bodies that started with separate core and mantle components, tracking the two types of particles to determine the result of the collision. They found that the fractional mass of the iron core of the largest post-collision remnant scaled with impact energy { (that is, the more energetic the impact, the higher the core fraction of the largest remnant)}, and lies between two limiting cases: model 1, in which cores always merge; and model 2, in which core material from the projectile is only accreted if the largest remnant is more massive than the { original} target (see \citealt{Marcus10}; \citetalias{Bonsor15}). As in \citetalias{Bonsor15} we assume that any { core material not accreted by the largest remnant is distributed to the smaller collision remnants (ensuring mass is conserved), starting by placing as much as possible in the second largest.}

In \citetalias{Bonsor15} the two empirical models from \citet{Marcus10} describing the mass of the iron core of the largest remnant of a collision were applied to each collision in the simulation in a post-processing step. This proved to be problematic as it could not account for the composition of the unresolved material\footnote{The simulations conducted for \citetalias{Bonsor15} had no compositional information at runtime, and so there was no composition for the unresolved material which was accreted continuously between collisions. The post-processing could determine the compositional change due to collisions, but the fraction of mantle or core material in the accreted unresolved material was unknowable.}; this had to be estimated in order to calculate embryo compositions. To avoid this we have incorporated the models used in \citetalias{Bonsor15} into our $N$-body code, which can now track the core mass of both resolved particles and debris annuli directly, applying the mantle stripping model as each collision occurs.

In \citetalias{Bonsor15} very little difference was found between the final core fractions of embryos using the two models from \citet{Marcus10}. We investigate the difference again in this work, and also define a third model as the mean of the original two,  $\mathrm{M}_{\mathrm{core, lr,3}} = (\mathrm{M}_{\mathrm{core, lr,1}} + \mathrm{M}_{\mathrm{core, lr,2}})/2$ (see Table \ref{t:coremodel}).
\begin{table*}
\centering
\caption{Mantle stripping models from \citet{Marcus10}. Model 3 is defined as the average of the original two: $\mathrm{M}_{\mathrm{core, lr,3}} = (\mathrm{M}_{\mathrm{core, lr,1}} + \mathrm{M}_{\mathrm{core, lr,2}})/2$.\label{t:coremodel}}
\begin{tabular}{l c c c}
\hline
Collision outcome &	M$_{\mathrm{core, lr}}$ Model 1 & 											M$_{\mathrm{core, lr}}$ Model 2  \\
\hline
Perfect merging &	M$_{\mathrm{core, targ}}$ + M$_{\mathrm{core, proj}}$ &	 						M$_{\mathrm{core, targ}}$ + M$_{\mathrm{core, proj}}$ \\
Partial accretion &	$\min$(M$_{\mathrm{lr}}$, M$_{\mathrm{core, targ}}$ +  M$_{\mathrm{core, proj}}$) &	M$_{\mathrm{core, targ}}$ + $\min$(M$_{\mathrm{core, proj}}$, M$_{\mathrm{lr}}$ - M$_{\mathrm{core,targ}}$) \\
Hit \& run &		M$_{\mathrm{core, targ}}$ &												M$_{\mathrm{core, targ}}$ \\
Erosive, supercatastrophic disruption &	$\min$(M$_{\mathrm{lr}}$, M$_{\mathrm{core, targ}}$ +  M$_{\mathrm{core, proj}}$) &	$\min$(M$_{\mathrm{lr}}$, M$_{\mathrm{core, targ}}$) \\
\hline
\end{tabular}
\end{table*}

{ Collisions are not the only process that can effect bulk composition and core-mantle ratio of planetesimals; the initial oxidation state may have varied across the terrestrial planet feeding zone, but we do not know what any such distribution would have been. For this reason, and so that we will see the effects due to collisions alone, we assume the same initial core fraction across the terrestrial planet region.
We explore two cases for this initial core fraction (both for planetesimals and unresolved debris) matching the values used in \citetalias{Bonsor15}: 0.35, representative of high iron content chondrites (EH), and 0.22, representative of low iron content chondrites (LL).}

Core and mantle material are always isolated in our model, and we do not allow any mixing or reequilibration to occur. Accreted core material is assumed to merge instantly with the core of the target since we expect planets to be (at least partially) molten during accretion. This is a significant simplification { for some proposed scenarios \citep[in which ongoing equilibration between metal and silicate alters the metal-silicate ratios, e.g.][]{Rubie15}}, but a reasonable one given that the timescales involved and the degree to which mixing would occur are still poorly understood \citep[e.g.][]{Dahl10}.

\subsection{The Grand Tack}

\citet{Walsh11} presented the Grand Tack model of the evolution of the Solar System in which the inner planetesimal disc is truncated by the inwards-then-outwards migration of Jupiter and Saturn. In this scenario, Jupiter forms first, creating a gap in the disc allowing it to migrate inwards via type II migration. Saturn grew more slowly and later began migrating inwards more rapidly than Jupiter, allowing Saturn to `catch up' with Jupiter, becoming trapped in the 2:3 mean motion resonance. This occurs as Jupiter reaches 1.5\,AU; the gaps in the gas disc opened by the gas giants then overlap, changing the balance of the torques, and causing the migration direction to reverse. Jupiter and Saturn then migrate outwards together as the gas disc dissipates, leaving the gas giants close to their present day orbits, and the outer Solar System in the appropriate configuration for a late instability (associated with the late heavy bombardment). Jupiter migrating inwards to 1.5\,AU before `tacking' truncates the inner disc at $\sim$1\,AU, producing the much sought after small Mars \citep{Walsh11,OBrien14}.

In the simulations that employ the Grand Tack we include Jupiter as an $N$-body particle and model Jupiter's migration by giving it an additional acceleration to match the velocity changes described by \citet{Walsh11}. Unlike previous Grand Tack simulations e.g.\ \citet{Walsh11}, \citet{OBrien14}, \citet{Rubie15}, which begin with { a possibly unrealistic bimodal distribution} of planetesimals and embryos and allow Jupiter to begin migrating at the start, we begin with a less evolved planetesimal disc, and allow embryos to begin growing via collisions while Jupiter remains at 3.5\,AU for several Myr before it begins migrating.

In this work we test the effect of Jupiter's inward then outward migration starting at two different times -- since Jupiter's formation time, and the gas disc lifetime are uncertain -- the commonly quoted time of ~3\,Myr (56k\,yr simulation time), and a `late' Grand Tack in which the migration begins ~9\,Myr (220\,kyr simulation time) after the birth of the solar nebula. We also test two timescales for migration, both the standard 100\,kyr (2.8\,kyr simulation time) value of \citet{Walsh11}, and a `slow' migration timescale of 600\,kyr (17\,kyr simulation time). { Using the expansion factor method makes matching the timescales problematic, we have opted to accelerate the migration by the same $f^2$ factor by which the accretion timescale is reduced. Comparison of the slow and standard timescales will show the effect of slower migration, but should also reveal any effect of the accelerated evolution relative to the migration caused by the expansion factor (the slow timescale is equivalent to the nominal migration rate accelerated only by a factor $f$).}

The acceleration, $\dot v_\mathrm{in(out)}$, given to Jupiter to model its inward (outward) migration is described by:
\begin{equation}
\dot v_\mathrm{in}  =  \frac{v_\mathrm{3.5 AU} - v_\mathrm{1.5 AU}}{\tau},
\end{equation}
\begin{equation}
\dot v_\mathrm{out}  =  \frac{v_\mathrm{1.5 AU} - v_\mathrm{5.2 AU}}{\tau} \, e^{(t - t_\mathrm{tack})/\tau},
\end{equation}
where $v_{\mathrm{1.5 AU}}$ is Jupiter's Keplerian velocity at 1.5\,AU -- the location at which it `tacks', $t$ is the time since the start of the simulation, $t_\mathrm{tack}$ is the time at which Jupiter's migration reverses direction, and $\tau$ is the migration timescale. Jupiter begins the simulation at 3.5\,AU, reaches an innermost semi-major axis of 1.5\,AU, and ends the simulation after the Grand Tack at $\sim$5.2\,AU. Jupiter feels the gravitational influence of the planetesimals just as any planetesimal or embryo, however, any object that collides with Jupiter is forced to merge.

We do not include Saturn nor the ice giants in this work as we expect them to have no significant effect on the terrestrial planet region, as discussed by \citet{Walsh11}. We also ignore the outer planetesimals originally situated beyond the orbit of Jupiter, which would be numerically very expensive to model at these resolutions, as it is not expected that a significant number would be scattered into the terrestrial region \citep[$\sim$3 percent of final planet mass,][]{OBrien14,Rubie15}, and most would be accreted after the end time of our simulations.

\subsection{Calm disc}

In the calm disc scenario the initial planetesimal disc extends from 0.5--1.5\,AU, with no giant planets included.

\subsection{Gas drag}

In this paper we present some simulations that include a gas density distribution within the protoplanetary disc (the effects of gas drag were not included in \citetalias{Bonsor15}). In these simulations we calculate the drag force on the planetesimals following the prescription of \citet{Adachi76} and \citet{Brasser07}.

The drag force is applied to each body in the simulation as an additional acceleration in the direction opposite to that of the body's relative motion through the gas,
\begin{equation}
\dot v_\mathrm{D}  =  - \frac{1}{2m} C_\mathrm{D} \pi r^2 \rho_\mathrm{g} ( \boldsymbol{v} - \boldsymbol{v}_\mathrm{g} )^2
\end{equation}
where $m$ and $r$ are the planetesimal mass and radius, $C_\mathrm{D}$ is the dimensionless drag coefficient, $\rho_\mathrm{g}$ is the gas density, and $\boldsymbol{v} - \boldsymbol{v}_\mathrm{g}$ is the velocity of the planetesimal relative to the gas velocity. The drag acceleration is calculated using the un-inflated planetesimal radius, and is applied only to resolved particles; unresolved material is assumed to accrete before its orbit is significantly affected by drag, or to be replenished by inward drift from larger radii.

The gas profile we adopt has one of two forms, either a power law matching the surface density of the solar nebula, or a third order polynomial obtained from a fit to the profile used by \citet{Walsh11}, that matches the results of hydrodynamic simulations from \citet{Morbi07}. The MMSN gas density is given by \citep{Brasser07},
\begin{equation}
\rho_\mathrm{g} = \rho_\mathrm{g,1} \, s^{-11/4} \, e^{-z^2/z_s^2},
\end{equation}
where $\rho_\mathrm{g,1}$ is the initial gas density at 1\,AU, $s$ is the distance from the central star in the $(x,y)$-plane, and $z_s = (0.047 \, s^{5/4})$ is the scale height of the gas disc, and the hydrodynamically derived gas density is given by,
\begin{equation}\label{hydrogas}
\begin{alignedat}{2}
&\rho_\mathrm{g} = \,& \frac{5.2}{s_{\mathrm{J}}} \left( 21880 \left( \frac{s}{s_{\mathrm{J}}} \right)^3 \, - \, 48010 \left( \frac{s}{s_{\mathrm{J}}} \right)^2 \right. \\
& & \left.  + \, 25420 \frac{s}{s_{\mathrm{J}}} \, + \, 663.3 \right) \frac{e^{-z^2/z_s^2}}{\sqrt{\pi} z_s},
\end{alignedat}
\end{equation}
if $s <= 0.9473 \, s_{\mathrm{J}}$, or,
\begin{equation}\label{hydrogas2}
\rho_\mathrm{g} =  \frac{5.2}{s_{\mathrm{J}}} \, 90 \, \frac{e^{-z^2/z_s^2}}{\sqrt{\pi} z_s},
\end{equation}
otherwise,
where $s_J$ is the instantaneous $s$ of Jupiter. These equations provide a good approximation for the gas profile given in \citet{Walsh11} interior to the orbit of Jupiter, the region with which we are concerned. We adopt a temperature profile $T = 280 \, s^{-1/2}$\,K as in \citet{Brasser07}, and thus the pressure gradient is given by $\eta = 2.13 \times 10^{-3} \, s^{1/2}$. The gas velocity is then obtained from,
\begin{equation}
\boldsymbol{v}_\mathrm{g} = \boldsymbol{v}_\mathrm{K} \, \sqrt{1-2\eta},
\end{equation}
where $\boldsymbol{v}_\mathrm{K}$ is the Keplerian velocity at a distance $s$ from the star.

In our Grand Tack simulations the gas density begins to decay exponentially as Jupiter migrates outwards through the disc, whereas in the calm disc simulations the gas density either decays over the same time period or remains constant throughout.

\newcounter{simnum}
\setcounter{simnum}{0}
\newcommand\simnumber{\refstepcounter{simnum}\thesimnum}
\begin{table*}
\centering
\caption{Summary of simulations presented in this work. \label{t:sims}}
\begin{tabular}{>{\simnumber}r r r r r r}
\hline
\multicolumn{1}{r}{Simulation} &	$N_i$ &		Initial core fraction &		Mantle stripping &	Gas disc &	Jupiter \\
\hline
\label{022f6nogas_l} &  			10\,000 &		0.22 &				1 &					None &			-- \\
\label{022f6nogasm2_l} &  		10\,000 &		0.22 &				2 &					None &			-- \\
\label{022f6nogasm3_l} &  		10\,000 &		0.22 &				3 &					None &			-- \\

\label{035f6nogas_l} &  			10\,000 &		0.35 &				1 &					None &			-- \\

\label{022f6nogas} &  			100\,000 &	0.22 &				3 &					None &			-- \\
\label{022f6nogas_2} &  			100\,000 &	0.22 &				3 &					None &			-- \\
\label{022f6nogas_3} &  			100\,000 &	0.22 &				3 &					None &			-- \\

\label{035f6nogas} &  			100\,000 &	0.35 &				3 &					None &			-- \\
\label{035f6nogasm1} &  			100\,000 &	0.35 &				1 &					None &			-- \\
\label{035f6nogasm2} &  			100\,000 &	0.35 &				2 &					None &			-- \\
\label{035f6nogas_2} &  			100\,000 &	0.35 &				3 &					None &			-- \\

\label{022f6simpgas_l} &  		10\,000 &		0.22 &				3 &					Constant MMSN&		-- \\
\label{035f6simpgas_l} &  		10\,000 &		0.35 &				3 &					Constant MMSN&		-- \\
\label{022f6simpgas_3} &  		100\,000 &	0.22 &				3 &					Constant MMSN&		-- \\
\label{035f6simpgas} &  			100\,000 &	0.35 &				3 &					Constant MMSN&		-- \\

\label{022f6simpgas_2_l}&  		10\,000 &		0.22 &				3 &					Dissipating MMSN&		-- \\
\label{035f6simpgas_2_l} &  		10\,000 &		0.35 &				3 &					Dissipating MMSN&		-- \\
\label{022f6simpgas_4} &  		100\,000 &	0.22 &				3 &					Dissipating MMSN&		-- \\
\label{035f6simpgas_2} &  		100\,000 &	0.35 &				3 &					Dissipating MMSN&		-- \\

\label{022GTJf6nogas_l} &  		10\,000 &		0.22 &				3 &					None &			Fast, early GT \\
\label{022GTJf6nogas} &  		100\,000 &	0.22 &				3 &					None &			Fast, early GT \\

\label{022GTJf6nogasf_l} &  		10\,000 &		0.22 &				3 &					None &			Slow, early GT \\
\label{022GTJf6nogasf} &  		100\,000 &	0.22 &				3 &					None &			Slow, early GT \\

\label{022GTJf6late_l} &  			10\,000 &		0.22 &				3 &					None &			Fast, late GT \\

\label{022GTJf6simpgas_l} &  		10\,000 &		0.22 &				3 &					MMSN &			Fast, early GT \\

\label{022GTJf6hydrogas_l} &  	10\,000 &		0.22 &				3 &					Hydrodynamic &	Fast, early GT \\
\label{022GTJf6hydrogas_2_l} &  	10\,000 &		0.22 &				3 &					Hydrodynamic &	Fast, early GT \\
\label{035GTJf6hydrogas_l} &  	10\,000 &		0.35 &				3 &					Hydrodynamic &	Fast, early GT \\

\label{022GTJf6hydrogasf_l} &  	10\,000 &		0.22 &				3 &					Hydrodynamic &	Slow, early GT \\

\label{022GTJf6latehg_l} &  		10\,000 &		0.22 &				3 &					Hydrodynamic &	Fast, late GT \\

\hline
\end{tabular}
\end{table*}
%

%%%%%%%%%%%%%%%%%%%% 			RESULTS 		%%%%%%%%%%%%%%%%%%%%

\section{Results}

The simulations discussed in the rest of this paper are summarised in Table \ref{t:sims}. The larger disc and added complexity required for the Grand Tack simulations leads to substantially longer running times, and so we were only able to conduct a small number of high resolution versions of these simulations. The similarity between the results of high and low resolution runs both for the calm disc and the Grand Tack model allows us to be confident that the low resolution simulations are sufficient to demonstrate the behaviour of the system.

\subsection{Planetary embryo evolution}

\begin{figure*}
\centering
\includegraphics[width=\textwidth]{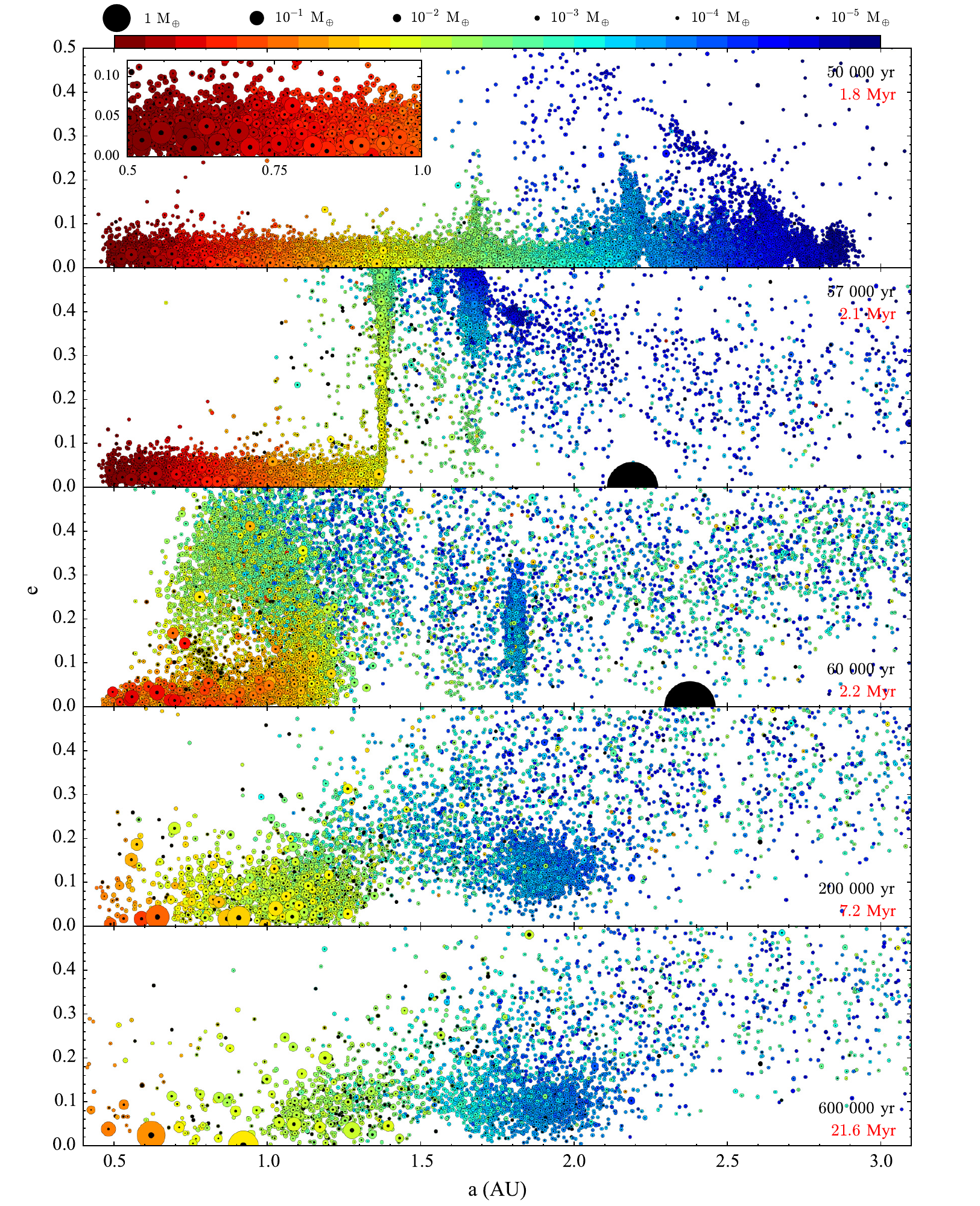}
\caption{Evolution of the eccentricity vs semi-major axis for growing terrestrial planet embryos in the Grand Tack model (simulation \ref{022GTJf6nogas}). The colours of the planetesimals represent the initial location of the material from which they are formed, as indicated by the colour bar. The sizes are proportional to the radii of the planetesimals, and the black centres represent their core mass fractions. The large black circle represents Jupiter, which is not to scale with the rest of the objects. Black and red labels show the simulation and effective timescales. The inset in the first panel shows a close-up of the inner part of the disc. (An animation of this figure is available online.)\label{f:a_e_GT}}
\end{figure*}

Figure \ref{f:a_e_GT} shows the semi-major axis vs eccentricity at five snapshots in time of a system of planetesimals evolving in the Grand Tack scenario. This broadly reflects the results of the Grand Tack as seen in other works, e.g.\ \citet{Walsh11}, \citet{OBrien14}, namely, Jupiter truncates the inner disc and scatters some material outwards, leaving some higher eccentricity planetesimals between 2--3\,AU to populate the asteroid belt. Note that the banded structure seen beyond $\sim$1.3\,AU in the third panel of Figure \ref{f:a_e_GT}, and the grouping of excited planetesimals seen at $\sim$1.9\,AU in the bottom two panels are caused by orbital resonances with Jupiter during its (outward) migration, and would likely be disrupted by the presence of Saturn.

\begin{figure*}
\centering
\includegraphics[width=\textwidth]{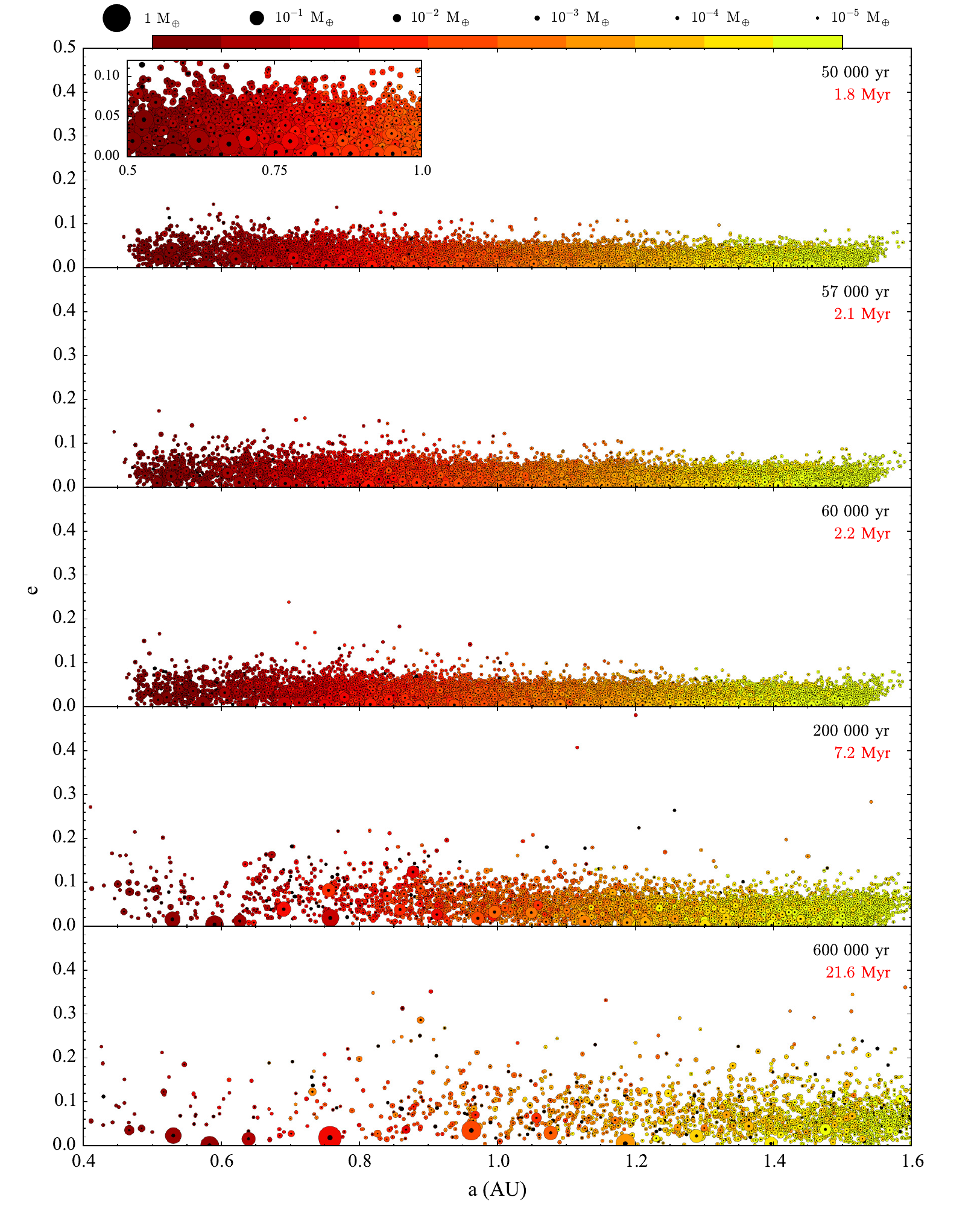}
\caption{Evolution of the eccentricity vs semi-major axis for growing terrestrial planet embryos in a calm protoplanetary disc (simulation \ref{022f6nogas}). The colours of the planetesimals represent the initial location of the material from which they are formed, as indicated by the colour bar. The sizes are proportional to the radii of the planetesimals, and the black centres represent their core mass fractions. Black and red labels show the simulation and effective timescales.  The inset in the first panel shows a close-up of the inner part of the disc. Note that the range of semi-major axis shown here is smaller than that in Figure \ref{f:a_e_GT}, but the colour scales for this region of the disc are the same. (An animation of this figure is available online.)\label{f:a_e_calm}}
\end{figure*}

Comparing the evolution under the Grand Tack model (Figure \ref{f:a_e_GT}) to the evolution of a calm disc environment (Figure \ref{f:a_e_calm}), it can be seen that planetesimal eccentricities become much higher in the Grand Tack scenario, and larger embryos grow more rapidly. Examining the colours of the planetesimals and embryos, which indicate the initial location of material, it is clear that some blue or green material originating beyond 1.5\,AU has been pushed into the inner disc by Jupiter's migration, and that the disc is more mixed compared to the calm disc scenario, with only orange, yellow or green objects present in the inner disc at the end of the Grand Tack simulation, compared to the calm disc in which the final colours of objects approximately match the initial colours at those locations. { It is worth noting that this mixing may be reduced compared to what would be seen with the nominal migration rate as the efficiency of resonant transport may have been changed by the $f^2$ increase in the migration rate.}


The results of our simulations for calm discs (simulations \ref{022f6nogas_l}--\ref{035f6simpgas_2}) are broadly similar to previous investigations e.g.\ \citet{Kokubo+Ida02,Leinhardt05}, with the same general differences noted in \citetalias{Bonsor15}, namely that the larger number of particles (smaller starting sizes) and the inclusion of a sophisticated collisional model with fragmentation increase the timescales for runaway and oligarchic growth, as would be expected. The main obvious difference, before we consider the planetesimal compositions, between the simulations presented here and those presented in \citetalias{Bonsor15} is the larger number of resolved fragments. This is due to the improved implementation of fragmentation described by \citet{Leinhardt15}.

\begin{figure*}
\centering
\begin{overpic}[width=\textwidth]{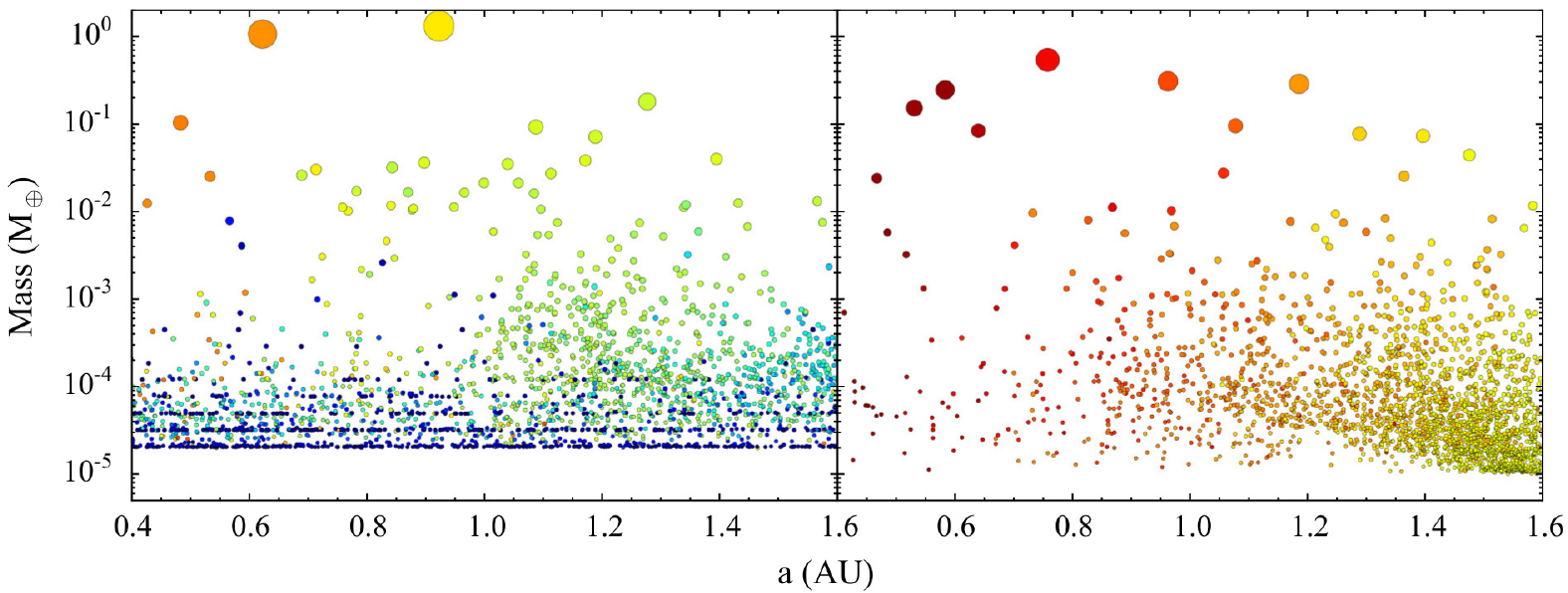}
\put(48,34){\normalsize GT}
\put(92,34){\normalsize Calm}
\end{overpic}
\caption{Mass distribution of planetesimals and embryos against semi-major axis at the end (600\,000\,yr simulation time) of a Grand Tack simulation (\ref{022GTJf6nogas}, left) and a calm disc simulation (\ref{022f6nogas}, right). The sizes of points are proportional to the radii of the planetesimals, and their colour indicates the mass-weighted radial source composition. (Animations of these figures are available online.)\label{f:a_mass}}
\end{figure*}

The mass distributions at the end of simulations \ref{022GTJf6nogas} and \ref{022f6nogas} are shown in { Figures \ref{f:a_mass} and \ref{f:massdist}}. Again this shows that some embryos grow larger in the same time period under the Grand Tack scenario. { Here we are seeing the effect of shepherding significant amounts of mass from beyond 1.5\,AU into the inner region of the disc; that the Grand Tack simulations result in embryos that are already one Earth-mass or more in only $\sim$20\,Myr of evolution is the result of this higher mass in the inner disc.} The distinct lines of dark blue coloured planetesimals seen in the left hand panel of Figure \ref{f:a_mass} show objects that underwent very little evolution in the outer part of the modelled disc (beyond $\sim$2.5\,AU) before being scattered onto highly eccentric orbits by Jupiter. They have subsequently undergone very little or no interaction with other planetesimals, but could be responsible for late delivery of volatile elements (e.g.\  H, C).

The right hand panel of Figure \ref{f:a_mass} shows a very similar result to previous work investigating calm protoplanetary discs (e.g.\ \citealt{Leinhardt15}; \citetalias{Bonsor15}); embryos grow and detach from the mass distribution of the planetesimal population. As noted by \citet{Leinhardt15} the embryos clean out the planetesimal population, starting in the inner disc where the evolution is faster, leading to an ``embryo front'' that propagates outwards through the disc, and there is no period where the total mass is split uniformly between planetesimals and embryos across the entire disc. This ``inside-out" growth that results from the more primitive and realistic initial condition is in contrast to many recent $N$-body models \citep[e.g.][]{Chambers13,Jacobson14}. This is also true in the Grand Tack scenario, but with the added complication of the influx of material shepherded by Jupiter.

\subsection{Core fractions}

\begin{figure*}
\centering
\begin{overpic}[width=\textwidth]{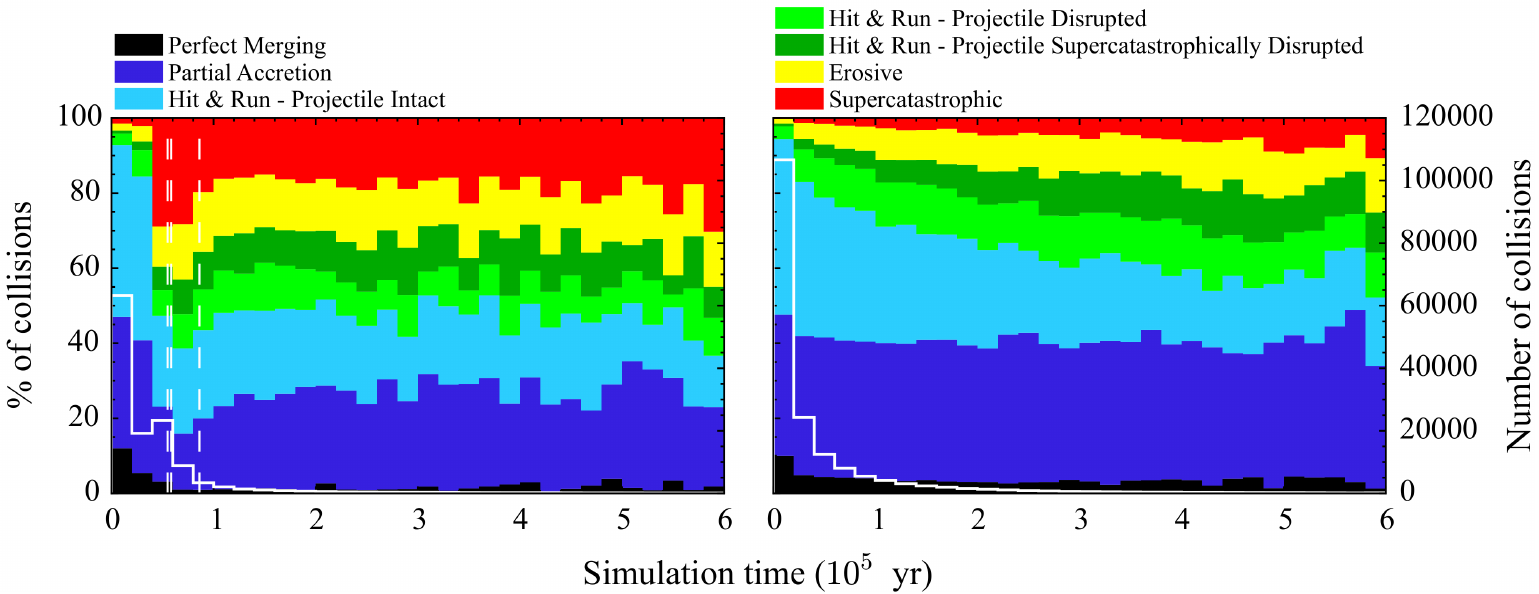}
\put(43,32){\normalsize GT}
\put(84,32){\normalsize Calm}
\end{overpic}
\caption{Collision history for a fast, early Grand Tack simulation (\ref{022GTJf6nogas}, left) and a calm disc simulation  (\ref{022f6nogas}, right) as a function of simulation time. The coloured histogram shows the percentage (left axes) of each type of collision occurring in each 20\,kyr time bin (simulation time). The white histogram shows the number of collisions that occurred in each time bin (right axes). The white dashed lines indicate the times of the start of Jupiter's migration, the reversal of the migration direction, and the approximate end of the outward migration (after ten outward migration timescales). Note that the first two of these lines are very close together on this scale, and that both occur in the third histogram bin.\label{f:coll_hist}}
\end{figure*}

Figure \ref{f:coll_hist} shows the fraction of each type of collision that occurred as a function of simulation time for the same Grand Tack simulation, \ref{022GTJf6nogas}, and calm disc simulation, \ref{022f6nogas} (both high resolution simulations that ignore gas drag, { the effects of gas drag will be discussed below}). It is immediately apparent that there are more erosive collisions (yellow) and supercatastrophic disruptions (red) under the Grand Tack scenario compared to the calm disc case. The left hand panel of Figure \ref{f:coll_hist} shows a spike in the number of collisions and fractions of supercatastrophic disruptions and erosive collisions during Jupiter's inward migration, and a higher fraction of these disruptive collision types than in the calm disc throughout the rest of the simulation.

The collision history for the calm disc simulation shown in the right hand panel of Figure \ref{f:coll_hist} is very similar to that shown in \citetalias{Bonsor15}. As was seen in both \citetalias{Bonsor15} and \citet{Leinhardt15} the fraction of disruptive collisions (yellow and red) increases with time as embryos grow, increasing the scattering of planetesimals. Note that the number of collisions decreases strongly with time in both scenarios (white lines), as the number of resolved particles falls. Again the fraction of accretional collisions is approximately constant with time.

\begin{figure*}
\centering
\begin{overpic}[width=\textwidth]{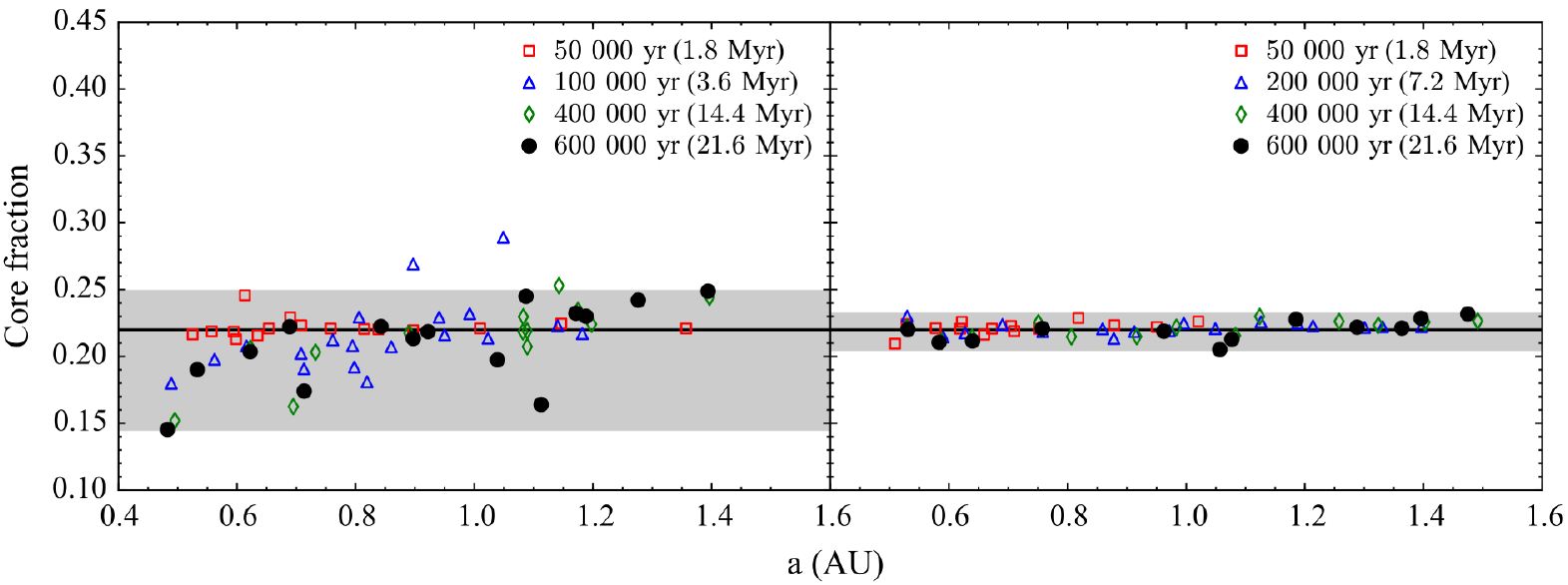}
\put(8,33){\normalsize GT}
\put(54,33){\normalsize Calm}
\end{overpic}
\caption{Core fractions of embryos at four times during a Grand Tack simulation (\ref{022GTJf6nogas}, left) and a calm disc simulation (\ref{022f6nogas}, right). Note that the red squares in the left hand panel represent a time before the migration begins. The grey area shows the final total range for the embryos in this region -- defined as objects with a mass of at least 2 lunar masses. In the simulations depicted in this figure the initial core fraction was set to 0.22, indicated by the solid line.\label{f:corefrac}}
\end{figure*}
Unsurprisingly the dynamical excitation of the Grand Tack leads to a much more violent environment. The significant numbers of planetesimals with very large black cores visible in Figure \ref{f:a_e_GT} immediately suggests that migration of Jupiter has had a major effect on the planetesimal core fractions. The influence of the migration on core fraction of the growing embryos is more clearly evident in Figure \ref{f:corefrac}. In this case with no gas and the standard fast, early migration (simulation \ref{022GTJf6nogas}, left panel of Figure \ref{f:corefrac}), there is a large variation in core fraction following Jupiter's migration (blue triangles), and some evolution towards lower core fractions is noticeable as the simulation proceeds (black circles). 
Even though the largest core fraction of an embryo falls during this later evolution, the maximum is still higher than is ever reached in the calm disc, and the range in embryo core fraction is substantially larger (0.15--0.25 compared to 0.20--0.23 for the calm disc).

\begin{figure*}
\centering
\begin{overpic}[width=\textwidth]{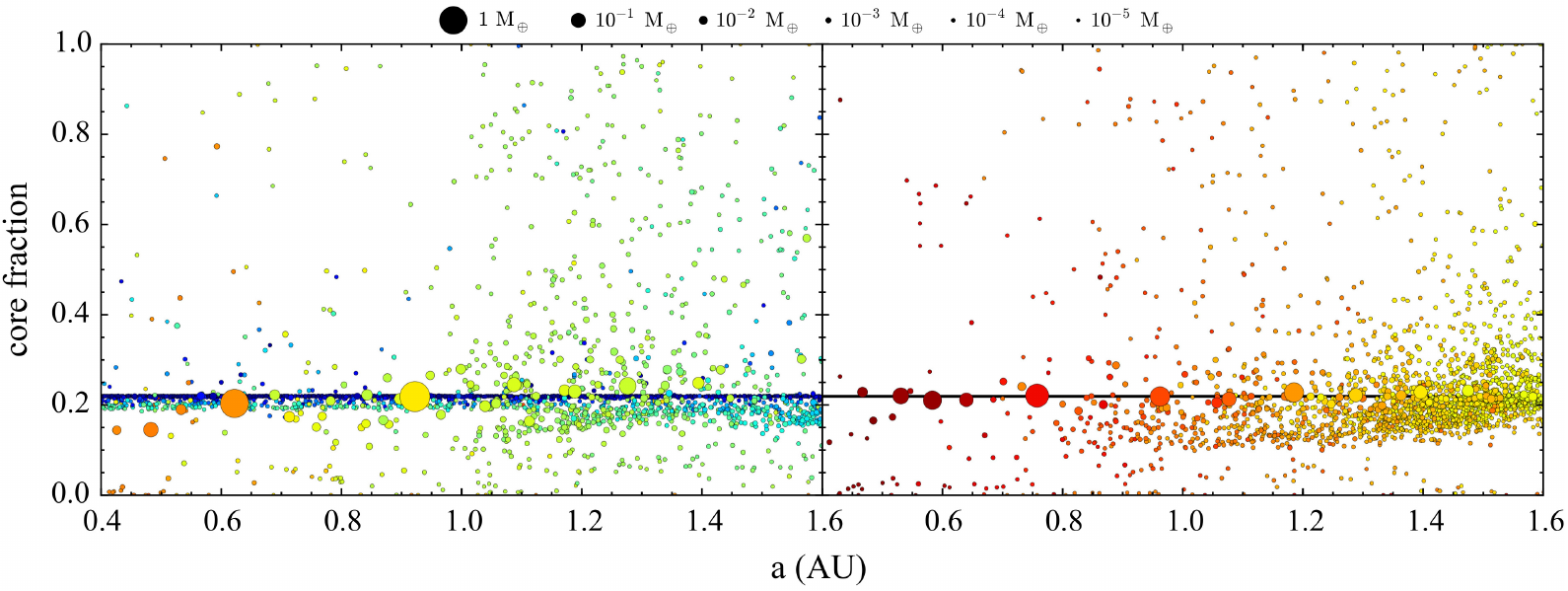}
\put(7,32){\normalsize GT}
\put(53,32){\normalsize Calm}
\end{overpic}
\caption{Core fractions of planetesimals and embryos at the end (600\,000\,yr simulation time) of a Grand Tack simulation (\ref{022GTJf6nogas}, left) and a calm disc simulation (\ref{022f6nogas}, right). The sizes of points are proportional to the radii of the planetesimals, and their colour indicates the mass-weighted radial source composition. The solid line indicates the initial core fraction. (Animations of these figures are available online.)\label{f:a_corefrac}}
\end{figure*}
Figure \ref{f:a_corefrac} shows the final core fractions of all resolved objects against their final semi-major axis, with the size of the points representing the radii of the planetesimals/embryos. In both simulations there is a small scatter in final core fractions of the embryos around the initial value of 0.22 (as seen in Figure \ref{f:corefrac}), { with the largest bodies in the left hand panel having core fractions very close to the initial value.} The smallest resolved planetesimals, however, show a huge variation in core fraction, with some planetesimals composed entirely of iron, and some entirely of silicate mantle material (again in the left hand panel there are many largely unevolved blue planetesimals with small semi-major axis on high eccentricity orbits that still have approximately their initial core fraction).

\subsubsection{The mantle stripping models}

We found, as in \citetalias{Bonsor15}, that the choice of mantle stripping law made little difference to the final core fractions of embryos. Since the original two models from \citet{Marcus10} should over and underestimate the core fraction of the largest remnant, all later simulations, and those discussed below, use the average model, model 3 (see Table \ref{t:sims}).

\subsubsection{Unresolved debris}\label{s:debris}

\begin{figure*}
\centering
\begin{overpic}[width=\textwidth]{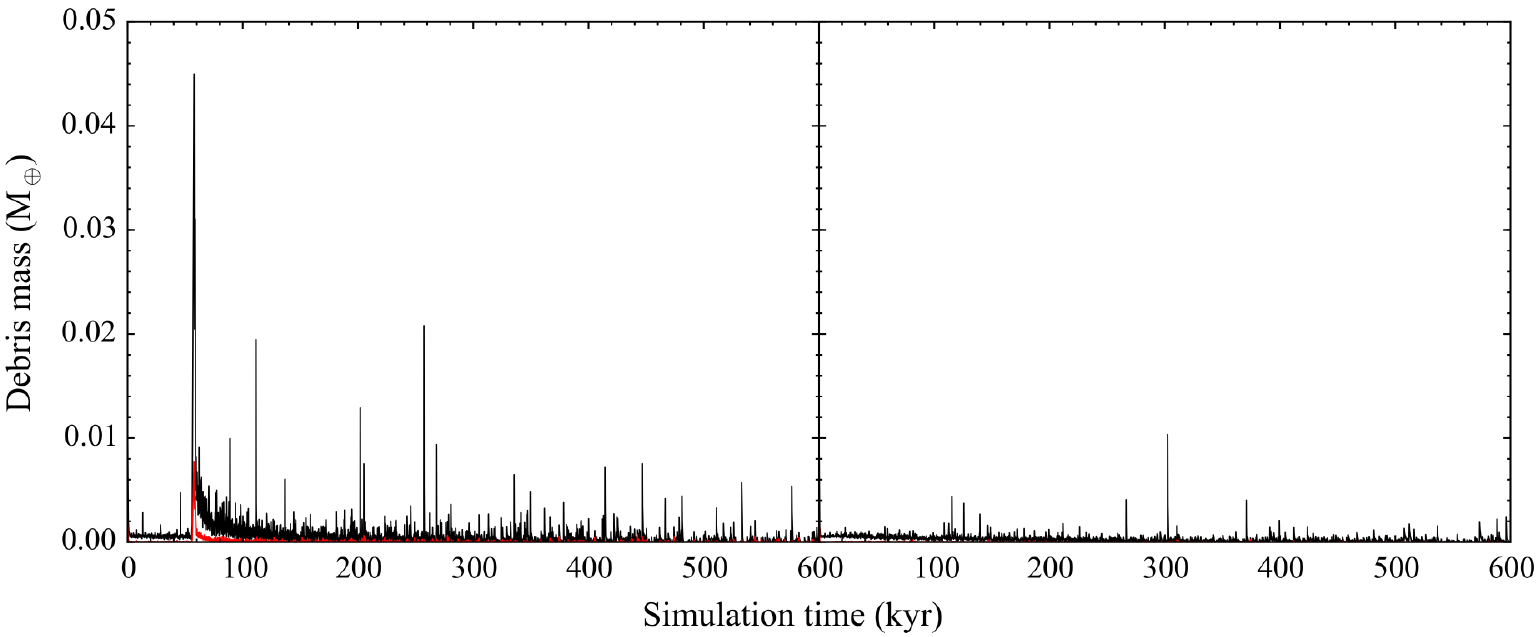}
\put(47.5,37){\normalsize GT}
\put(91,37){\normalsize Calm}
\end{overpic}
\caption{Unresolved debris mass as a function of simulation time for a Grand Tack simulation (\ref{022GTJf6nogas}, left) and a calm disc simulation (\ref{022f6nogas}, right). The black line indicates the total mass in unresolved debris between 0.5 and 1.5\,AU, and the red line indicates the corresponding mass of core material.\label{f:debris}}
\end{figure*}
Figure \ref{f:debris} shows the mass of unresolved debris present in the inner disc (0.5--1.5\,AU) as the simulations progress. There is a clear spike in the left panel of Figure \ref{f:debris} corresponding to Jupiter's migration, and a number of smaller spikes in both simulations caused by collisions.

In both simulations shown in Figure \ref{f:debris} the initial core fraction of all planetesimals and the initial unresolved material is 0.22; the average core mass fraction of the debris after 100\,kyr is 0.14 and 0.10 in the Grand Tack and calm disc simulations respectively, and the average mass in unresolved debris after 100\,kyr is 4$\times$10$^{-4}$\,M$_\oplus$ and 2$\times$10$^{-4}$\,M$_\oplus$. The unresolved material is enriched in silicate mantle material, which is as we would expect since it is mantle that is lost first in fragmenting collisions. As was shown by \citep{Leinhardt15}, the total mass in the unresolved component is always a small fraction of the total system mass, and never builds to a significant mass.

\subsection{Migration timescales}

\begin{figure*}
\centering
\begin{overpic}[width=0.485\textwidth]{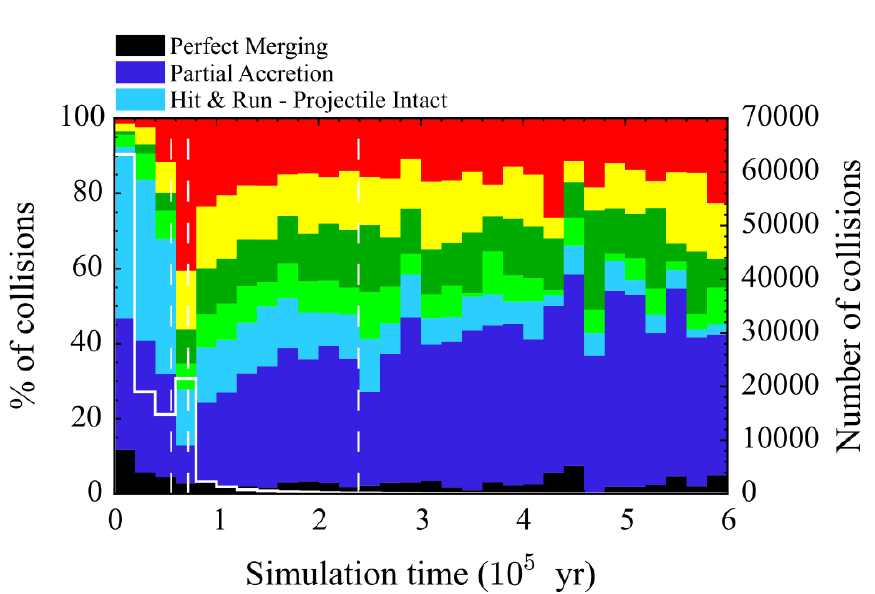}
\put(82,63){\normalsize slow GT}
\end{overpic}
\begin{overpic}[width=0.485\textwidth]{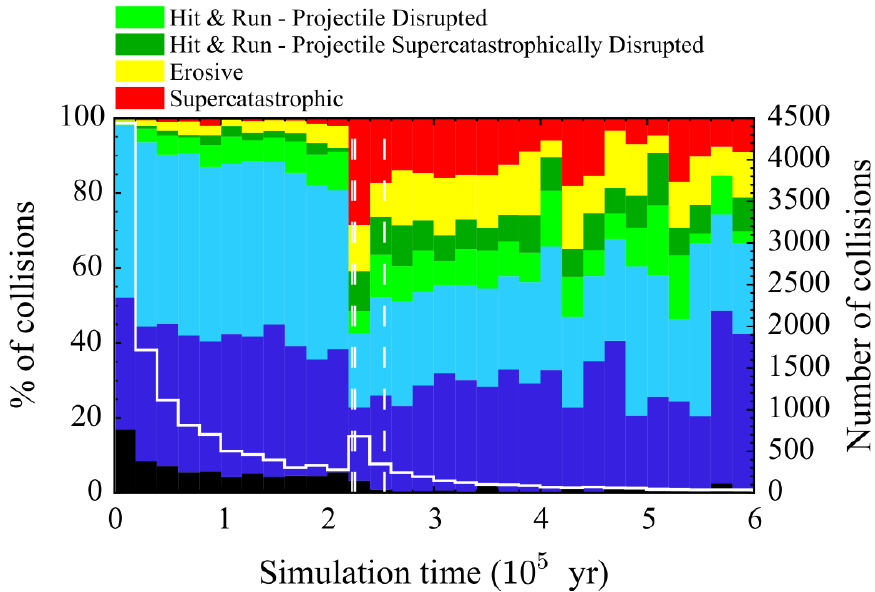}
\put(82,63){\normalsize late GT}
\end{overpic}
\caption{Collision history of Grand Tack simulations as a function of simulation time, for a high resolution simulation with slow, early migration (\ref{022GTJf6nogasf}, left), and a low resolution simulation with fast, late migration (\ref{022GTJf6late_l}, right). The coloured histogram shows the percentage (left axes) of each type of collision occurring in each 20\,kyr time bin (simulation time). The white histogram shows the number of collisions that occurred in each time bin (right axes). The white dashed lines indicate the times of the start of Jupiter's migration, the reversal of the migration direction, and the approximate end of the outward migration (after ten outward migration timescales). Note that low resolution simulations tend to have more perfect hit and run collisions due to the smaller range of particle sizes.\label{f:coll_hist_GT}}
\end{figure*}

\begin{figure*}
\centering
\begin{overpic}[width=\textwidth]{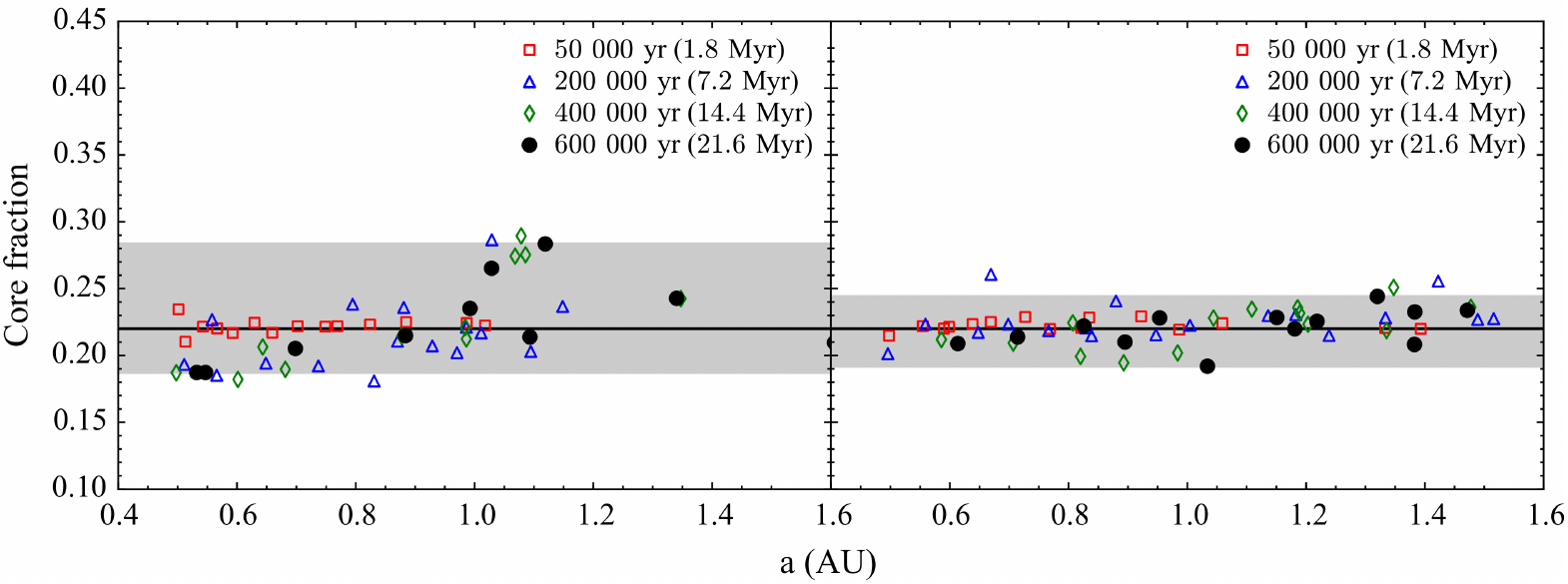}
\put(8.5,33){\normalsize slow GT}
\put(54,33){\normalsize late GT}
\end{overpic}
\caption{Evolution of the core fractions of embryos during Grand Tack simulations for a high resolution simulation with slow, early migration (\ref{022GTJf6nogasf}, left), and a low resolution simulation with fast, late migration (\ref{022GTJf6late_l}, right). The grey area shows the final total range for the embryos in this region -- defined as objects with a mass of at least 2 lunar masses. In the simulations depicted in this figure the initial core fraction was set to 0.22, indicated by the solid line. (Animations of these figures are available online.)\label{f:corefracGT}}
\end{figure*}

Figures \ref{f:coll_hist_GT} and \ref{f:corefracGT} show the collision histories and embryo core fractions resulting from slower migration (simulation \ref{022GTJf6nogasf}) and late migration (simulation \ref{022GTJf6late_l}).

The left hand panel of Figure \ref{f:coll_hist_GT} shows that the slower migration leads to a larger spike in the fraction of disruptive collisions (yellow and red), and a decrease in the fraction of `perfect' hit and run collisions (light blue) compared to the standard Grand Tack scenario (left panel of Figure \ref{f:coll_hist}). The hit and run collisions in which the projectile remains intact (light blue) cannot affect the composition of either body in our model, whereas the partial accretion collisions (dark blue) that replace them in this slower migration scenario can. The slower migration leads to a noticeable increase in the maximum final core fraction (0.26--0.28 for slow migration compared to 0.24--0.25 for the fast scenario), as shown in the left panel of Figure \ref{f:corefracGT} compared to the left panel of Figure \ref{f:corefrac}.

The late migration (shown in the right panels of Figures \ref{f:coll_hist_GT} and \ref{f:corefracGT}) causes an increase in the fraction of disruptive collisions similar to the standard
 fast, early migration simulations, but the embryo core fractions in this much more evolved planetesimal system do not undergo the same changes. The embryos have grown large enough during their longer period of evolution that the migration has little effect on their composition. The variation in the final core fractions is significantly smaller than in the early Grand Tack simulations, and there is no large variation after Jupiter's migration (green diamonds) to be averaged out by the end of the simulation (black circles).

\subsection{The effects of gas drag}

\subsubsection{Calm disc}

\begin{figure*}
\centering
\begin{overpic}[width=\textwidth]{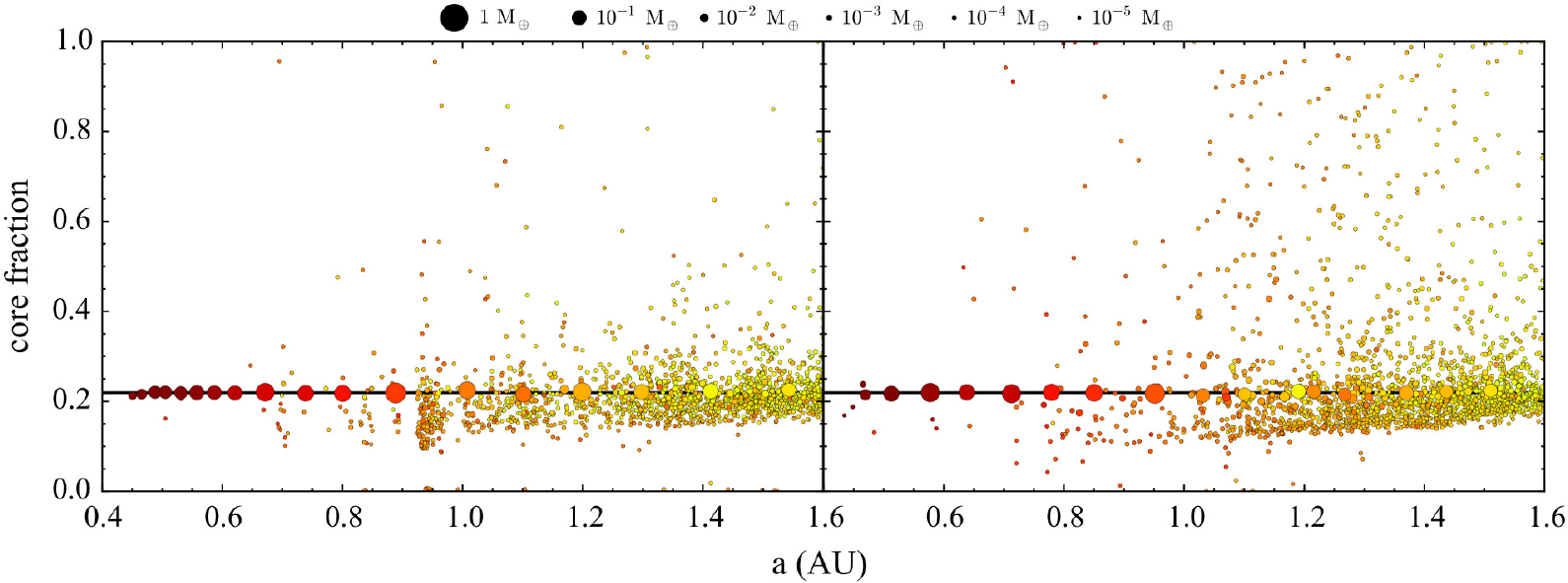}
\put(7,32){\normalsize const. gas}
\put(53,32){\normalsize diss. gas}
\end{overpic}
\caption{Core fractions of planetesimals and embryos at the end (600\,000\,yr simulation time) of calm disc simulations with constant gas (\ref{022f6simpgas_3}, left), and dissipating gas (\ref{022f6simpgas_4}, right). The sizes of points are proportional to the radii of the planetesimals, and their colour indicates the mass-weighted radial source composition. The solid line indicates the initial core fraction. (Animations of these figures are available online.)\label{f:acf_gas}}
\end{figure*}

\begin{figure*}
\centering
\begin{overpic}[width=\textwidth]{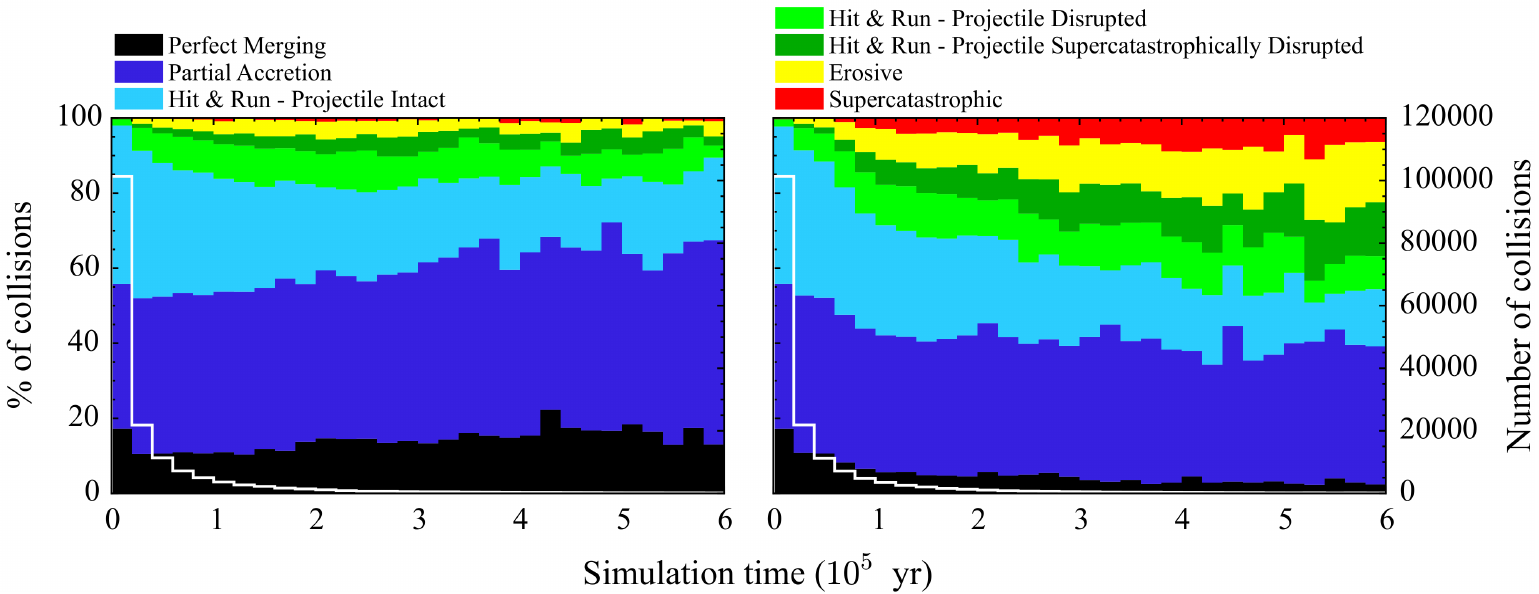}
\put(37,32){\normalsize const. gas}
\put(81,32){\normalsize diss. gas}
\end{overpic}
\caption{Collision history for high resolution calm disc simulations with constant gas (\ref{022f6simpgas_3}, left), and dissipating gas (simulation \ref{022f6simpgas_4}, right), as a function of simulation time. The coloured histogram shows the percentage (left axes) of each type of collision occurring in each 20\,kyr time bin (simulation time). The white histogram shows the number of collisions that occurred in each time bin (right axes). \label{f:coll_hist_gas}}
\end{figure*}

We have tested the influence of gas drag from a MMSN gas disc on the evolution of a calm planetesimal disc under two scenarios. In the first, the gas has a density that is constant in time throughout the duration of the simulations; in the second the gas begins to dissipate after 2.1\,Myr (58\,kyr simulation time) -- approximately 3\,Myr after the birth of the solar nebula -- the surface density then decays exponentially on a timescale of 100\,kyr (2.8\,kyr simulation time). These timescales were chosen to match those we use in the Grand Tack simulations (see section \ref{s:GTgas}), which match those from \citet{Walsh11}.

The gas lifetime in the first scenario is probably unrealistically long compared to observed gas disc lifetimes \citep[$\sim$6\,Myr,][]{Haisch01}, but provides a useful insight into the effects of gas drag on disc and planetary embryo evolution.

This scenario produces smaller oligarchs compared to the case without gas. Such an effect of gas drag was discussed by \citet{Tanaka97}, it slows the growth of embryos as planetesimals become trapped outside the embryo feeding zones. The left hand panel of Figure \ref{f:acf_gas} shows that the inner region of the disc has been cleared to a greater degree than in simulations without gas drag (right panel of Figure \ref{f:a_corefrac}). This reflects the more rapid evolution of the system due to gas drag, as seen by \citet{Wetherill93} and \citet{Kokubo00}.

Embryos grow more quickly in the runaway growth phase due to the eccentricity damping caused by the gas, however, during the oligarchic growth phase the embryos do not grow to be as massive as they are unable to sweep up the planetesimals. We are thus left at the end of our simulations with a system that has features that are indicative of both more and less evolved systems than the case without gas.

Figure \ref{f:coll_hist_gas} shows the significant change in the fraction of collision types when gas is included; the left hand panel shows that gas drag results in significantly fewer erosive and disruptive collisions (yellow and red), and a greater fraction of perfect merging collisions (black) compared to the case without gas drag (Figure \ref{f:coll_hist}, right). Gas drag damps the eccentricities and inclinations of the planetesimals, resulting in lower average collision velocities. The lack of these erosive and disruptive collisions in the constant gas scenario results in very few collisions that can significantly affect the core-mantle balance of the growing embryos. The final distribution of core fractions (left panel of Figure \ref{f:acf_gas}) therefore shows very little change from the initial value, and much less variation than the simulations without gas drag (Figure \ref{f:a_corefrac}, right panel).

The second scenario, in which the gas dissipates early in the simulation, unsurprisingly has similarities to both the constant gas scenario and the `no gas drag' scenario. The gas drag again facilitates the rapid growth of small embryos, however, once the gas density begins to decay the embryo eccentricities can be excited, allowing them to collide, producing a smaller number of larger embryos that go on to continue clearing the planetesimal population.

The collision histogram (right panel of Figure \ref{f:coll_hist_gas}) looks similar to the no gas drag case (Figure \ref{f:coll_hist}, right panel) once the gas begins to dissipate, but in the first 58\,kyr (2.1\,Myr effective time) -- during which the majority of the collisions occur -- there are fewer erosive events and more perfect merging collisions. This likely leads to the smaller variation in final core fraction shown in Figure \ref{f:acf_gas} (right hand panel) for this dissipating gas scenario compared to the case without gas drag (Figure \ref{f:a_corefrac}, right panel).

\subsubsection{The Grand Tack scenario} \label{s:GTgas}

Including gas in the standard fast, early migration scenario leads to embryo core fractions that evolve in a similar way to the simulations with slower migration, with the larger variation and enhanced final core fractions compared to the case without gas drag (see Figure \ref{f:all_cf_range}). Whilst the collision histograms look most similar to the fast, early migration without gas, the inclusion of gas drag has prevented the enhanced core fractions from being reduced by the end of the simulation. The results are similar for both a MMSN gas density profile (simulation \ref{022GTJf6simpgas_l}) and the hydrodynamic derived gas density profile (simulation \ref{022GTJf6hydrogas_l})  used by \citet{Walsh11}.

Including gas drag in the Grand Tack scenario increases the maximum core fraction compared to the scenario without gas drag; this is the opposite effect to that in the calm disc scenario, in which the damping of eccentricities caused by the drag force leads to a smaller variation in core fraction. The excitation caused by Jupiter's migration has a substantially larger effect than the damping caused by the drag force in a MMSN gas disc.

%%%%%%%%%%%%%%%%%%%% 			DISCUSSION 		%%%%%%%%%%%%%%%%%%%%

\section{Discussion}

\begin{figure}
\centering
\includegraphics[width=0.48\textwidth]{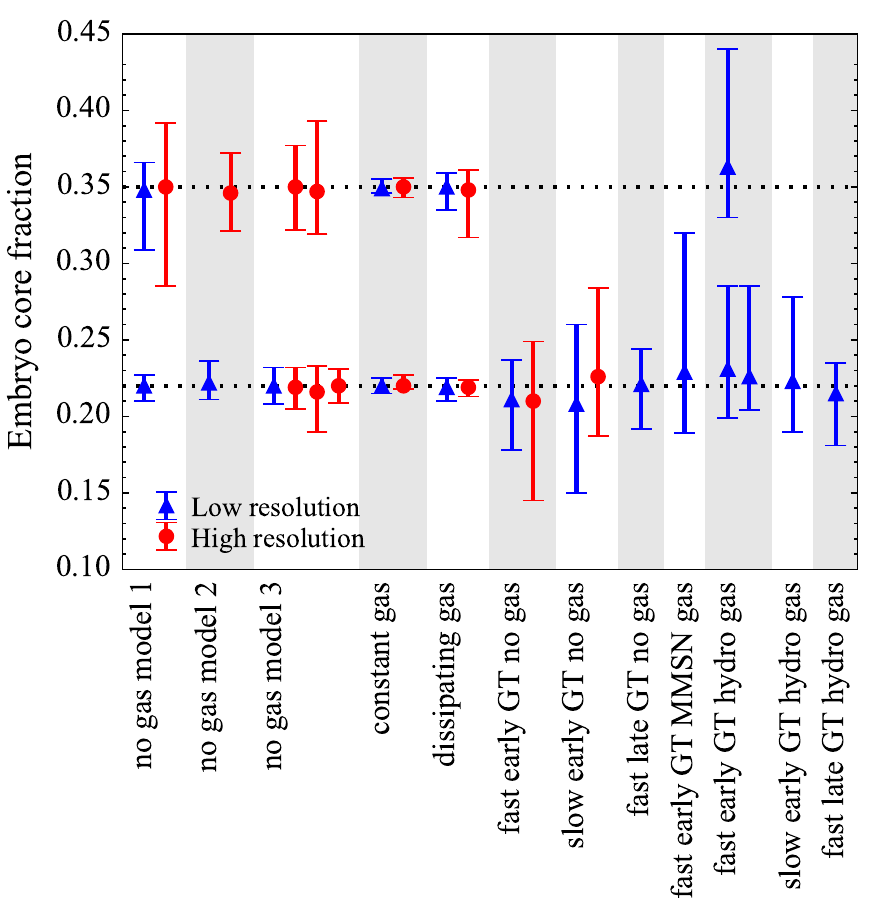}
\caption{Mean, minimum and maximum final embryo core mass fractions from all simulations. The dotted lines indicate the initial core fractions. Model 1, 2 and 3 refer to the mantle stripping laws.\label{f:all_cf_range}}
\end{figure}
The mean and range of embryo core mass fractions from all of the simulations are shown in Figure \ref{f:all_cf_range} (where we define embryos as bodies with a mass of at least 2 lunar masses; note that this mean value is not weighted by the embryo mass, and that both core material and mantle material are conserved in these simulations). It is important to note that in all cases we find both embryos that show an enhanced core fraction, and embryos that show a decreased core fraction.

It is clear from Figure \ref{f:all_cf_range} that the dynamical excitation caused by Jupiter's Grand Tack leads to a higher maximum and larger range of embryo core fractions compared to the calm scenario. This is not entirely surprising, Jupiter's Grand Tack excites the orbits of many planetesimals and embryos, which makes them more likely to undergo more energetic collisions. This leads to larger numbers of disruptive collisions, and it is these disruptive collisions that can strip off large amounts of mantle, and even excavate material from the cores of planetesimals or embryos, allowing core material to be redistributed to other objects.

{ \citet{Dwyer15} investigated the compositional evolution of terrestrial planets during the late stages of accretion by post-processing low $N$ simulations of the giant impact phase. In this work we have studied an earlier stage of planet formation: the runaway and oligarchic growth phases, so a direct comparison is not possible. \citeauthor{Dwyer15} determined the core fractions of bodies resulting from imperfect collisions using a similar approach to our model 1, however, rather than using the unresolved debris method, the simulations they analysed either placed all the fragment mass into equal mass bodies, or, if this mass was below the resolution limit, it was simply accreted by the target (or projectile in the case of hit and run collisions,  \citealt{Chambers13}). This means that a significant amount of mantle had to be stripped before a change in composition would occur. They found a greater scatter in silicate mass fraction for lower mass bodies; similarly we find a greater variation in composition for smaller bodies, though it should be noted that we cover an earlier stage of evolution here, and that our mass limit is more than 100 times smaller. In the Grand Tack model we find some embryos with masses approaching Earth-mass (although their evolution is not complete), which tend to show only a small variation in core fraction similar to the results of the calm disc simulations (e.g.\ Figure \ref{f:a_corefrac}). The smaller embryos in the Grand Tack generally exhibit a larger range of core fractions, though these are still modest in comparison to the range shown by the smallest remaining planetesimals in both scenarios.}

The range of final embryo core fractions in the calm scenario is smaller than in \citetalias{Bonsor15}, which was by necessity an uncertain estimate due to the way in which accretion of unresolved material was handled. The updated range of final core fractions fit within the ranges from \citetalias{Bonsor15}, but have more conservative maximum values. { Even with our direct tracking of the composition of the unresolved debris there are still several limitations with this method. The dynamics of this material and any collisional evolution are ignored. In a real system a small fraction of this mass would likely be ground down to dust and lost via Poynting-Robertson drag, and the composition of individual sub-200-km bodies would also evolve. Since the unresolved material is enhanced in silicate compared to the initial conditions, any loss of this material from the system would further enhance core fractions of the resolved bodies. However, most of the mass in the unresolved debris would be in bodies with masses just below the resolution limit, only a tiny fraction of the mass would be in the form of dust that could be lost, and so we expect the effects of this to be minimal.} The updated results for the calm scenario lead to smaller shifts in composition than seen in \citetalias{Bonsor15}, but the collisional evolution of chondritic planetesimals could still produce embryos with Earth like composition in extreme cases with high initial core fractions.

As has been noted, we see a much larger variation in core fraction in the Grand Tack models than in the calm scenario. The fact that the maximum is higher when Jupiter's migration is slower, or when we include the effects of gas drag, is more difficult to understand. Figure \ref{f:coll_hist_GT} shows that there are fewer collisions after the Grand Tack in these cases compared to the scenario without gas drag, the slightly more evolved systems then undergo fewer collisions that could work to reduce the variation in core fraction that is present immediately after Jupiter's migration in all four cases (Figures \ref{f:corefrac} and \ref{f:corefracGT}). { In the case of slower migration, it is expected that the efficiency of resonant transport will be increased, which would lead to a larger number of excited planetesimals and hence a greater number of erosive collisions in the inner regions of the disc. Gas drag will cause the smallest fragments produced via collisions during Jupiter's inward migration to lose eccentricity and move inwards towards the Sun, and since these fragments will be dominated by mantle material, gas drag can act to enhance the core fractions of embryos closest to the tack point.} On the other hand, if the Grand Tack occurs late, too much evolution has already occurred, and the excitation is not sufficient to cause large variation in the core fractions of the now larger embryos. This suggests that the exact composition of the Earth could be sensitive to the timing of Jupiter's migration under the Grand Tack scenario, and that with a very good understanding of the evolution of the Earth's composition, it may be possible to `date' the Grand Tack.

Figures \ref{f:corefrac} and \ref{f:corefracGT} reveal a noticeable gradient in core fraction across the region occupied by the terrestrial planets, with lower core mass fractions interior to $\sim$1\,AU, and higher core fractions beyond this point. It is the region beyond $\sim$1\,AU that is most affected by the migration of Jupiter, with much of the material originally in this region excited onto high eccentricity orbits before relaxing as Jupiter migrates outwards again (see Figure \ref{f:a_e_GT}). { This causes some silicate rich fragments to be scattered out of this region (both inwards and outwards), leaving the disc enriched in iron between 1--1.5\,AU, and silicate enriched interior to 1\,AU. This is presumably sensitive to the location of the innermost point of Jupiter's migration, which was specifically chosen by \citet{Walsh11} to affect the disc outside 1\,AU. It is also important to remember that we have used a uniform initial core fraction, and it is expected that the planetesimals in a real system would have some variation in initial core fraction \citep[e.g.][]{Righter06}. Our final solar system clearly shows that the effects of the giant impact stage are significant and stochastic, with the high Fe/Mg ratio for Mercury possibly related to a giant impact and the preserved embryonic Fe/Mg for Mars.
}

\subsection{Mixing and volatile delivery}

\begin{figure*}
\centering
\begin{overpic}[width=1.0\textwidth,trim={0 1.2cm 0 0},clip]{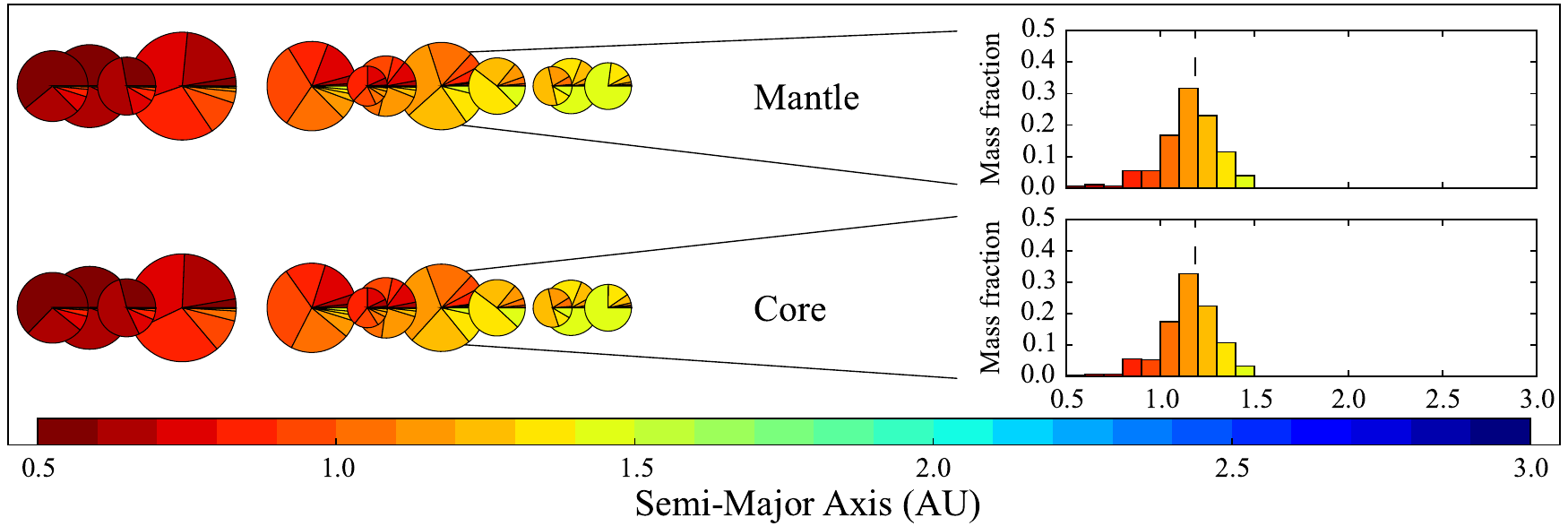}
\put(91,22){\normalsize Calm}
\end{overpic}\\
\begin{overpic}[width=1.0\textwidth]{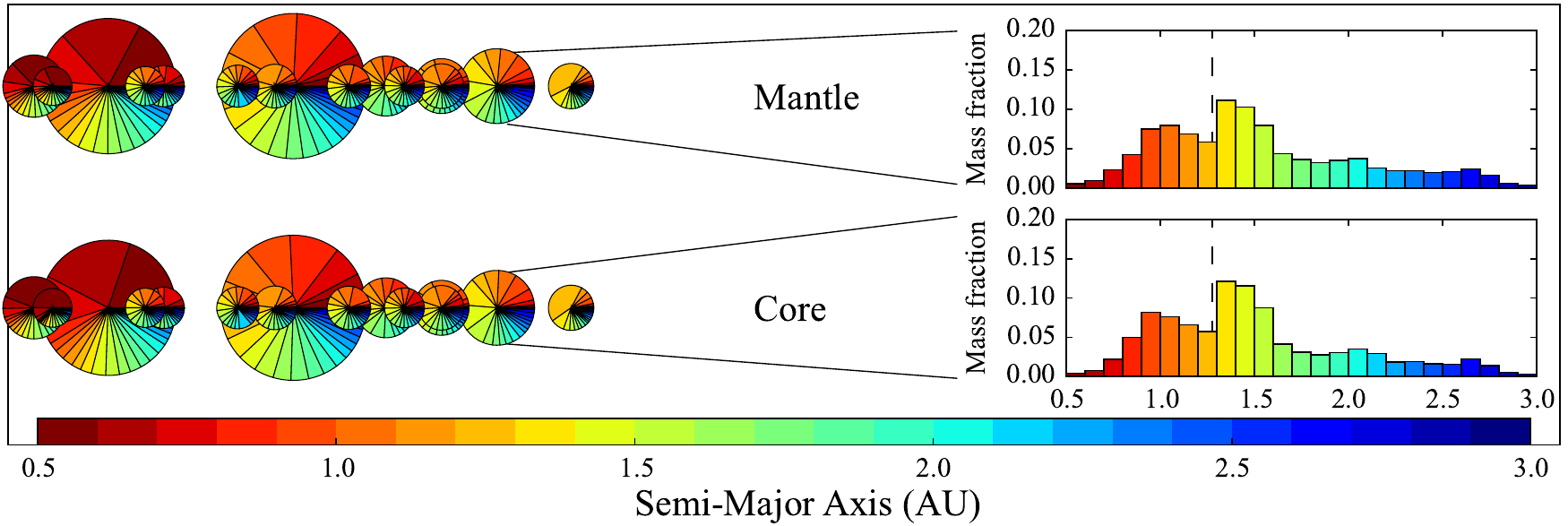}
\put(93,29){\normalsize GT}
\end{overpic}\\
\caption{Composition pies for each planetary embryo at the end point of a calm disc simulation without gas drag (\ref{022f6nogas}, upper), and a Grand Tack simulation with fast, early migration (\ref{022GTJf6nogas}, lower). The embryos are shown at their instantaneous semi-major axis, with the size of each pie proportional to the object's radius. Each sector of a pie shows the mass fraction of the embryo's mantle or core originating from the region indicated by the colour bar. The histograms on the right show this breakdown for the indicated embryo, the dashed line showing its location. \label{f:mixingGT}}
\end{figure*}

The `origin histogram' (see Figure \ref{f:mixingGT}) of each particle in our simulations tracks the initial locations of the mass of which that particle is comprised, so we can easily examine our final embryos and determine from which regions of the disc they have accreted planetesimals. The first panel of Figure \ref{f:mixingGT} shows the composition in terms of radial location of embryos produced in calm disc simulations. This shows a very similar result to that found by \citet{Leinhardt15}, that there is some localised mixing with most of the mass accreted by each embryo originating from the regions 0.1\,AU either side of its position at the end of the simulation.

Figure \ref{f:mixingGT} (lower panel) shows the mixing in the same way for the Grand Tack model. It is clear that Jupiter's migration in this model causes a much greater degree of mixing, with almost all embryos containing some significant fraction of material originating beyond 2\,AU, and no pair of adjacent 0.1\,AU regions dominating the mass of any embryo that remains in the terrestrial planet region. Figure \ref{f:mixingGT} also suggests a much more uniform radial composition due to this increased mixing than is seen in any of the calm disc simulations. Again we note that the accelerated migration may have reduced the efficiency of transport and thus caused reduced mixing.

In this work we have assumed a uniform initial core fraction across the planetesimal disc, this allows for a clearer investigation of the effects of collisional evolution, but in a real system we would expect some variation in oxidation state with heliocentric distance \citep[e.g.][]{Righter06}. This would be expected to have some effect on the final core fractions of the terrestrial planets, especially in the calm scenario which shows little mixing. The much more mixed embryos produced in the Grand Tack simulations may not be affected significantly by this simplification, however, the effect on the resulting system of planets will depend on the nature of the initial variation.

In the Grand Tack model all embryos acquire similar fractions of `blue' material (3--8 percent of final mass from beyond 2.5\,AU, see Figure \ref{f:mixingGT}), that is often assumed to contain more water than planetesimals formed closer to the Sun \citep[e.g.][]{Raymond09}. If planetesimals originating beyond 2.5\,AU are 10 percent water by mass (as is commonly assumed in similar calculations, e.g.\ \citealt{OBrien06}), the embryos at the end of these simulations would be 0.4--0.8 percent water, and the larger embryos would already have acquired enough to match the Earth's water content. 
Here we have used only planetesimals originating interior to the orbit of Jupiter, \citet{OBrien14} similarly estimate that any Earth mass planet forming under the Grand Tack would acquire the required amount of water from planetesimals originating beyond the orbit of Jupiter (which we have neglected). 
However, it is uncertain what fraction of the volatiles from planetesimals accreted during this period would be retained by the final planet, and water from both populations of planetesimals may be required if a substantial fraction is lost due to collisions.

\subsection{The fate of planetesimals}

\citet{Bottke06} proposed that the iron meteorites, which originate today from the inner asteroid belt, are remnants of planetesimals formed in the terrestrial planet region rather than fragments of bodies formed in the main belt. One outstanding question from \citeauthor{Bottke06}'s work is whether differentiated planetesimals from the terrestrial planet region could actually produce these iron-rich bodies via disruption. The results of this work suggest that many iron-rich, and complimentary silicate-rich, fragments are indeed produced by the collisional evolution of the inner disc, as seen in Figure \ref{f:a_corefrac}. Though we see very few of these remnant planetesimals scattered into the asteroid belt region, it is important to note that the initial distribution of planetesimals in our calm disc simulations does not extend beyond 1.5\,AU, and that such material may be important in scattering fragments into the main belt region (and keeping them there).

It has recently been realised that many white dwarfs display extrasolar planetary debris in their spectra \citep[e.g.][]{Jura14}. This material shows a diversity in the iron contents of accreted planetesimal and small embryo sized bodies \citep[e.g.][]{Farihi11,Gaensicke12,Raddi15} that could represent leftover collisionally processed planetesimals similar to those produced in our simulations (e.g. Figure \ref{f:a_corefrac}). The presence of differentiated rocky debris in white dwarf atmospheres appears to be a common phenomenon, which may suggest that many extrasolar planetary systems undergo similar collisional evolution to that shown in this work.

\subsection{What does this mean for the bulk composition of planetary embryos?}

\begin{figure}
\centering
\includegraphics[width=0.48\textwidth]{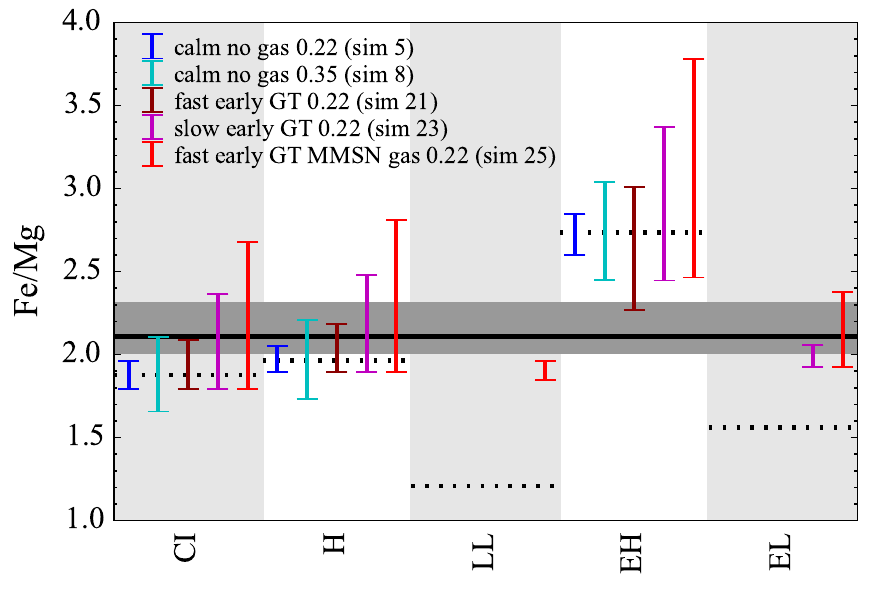}
\caption{{ Bulk Fe/Mg ratios for collisionally processed embryos. Using the range of final core fractions achieved in each simulation, the vertical bars show the range of possible Fe/Mg in embryos for a series of scenarios starting with different initial chondritic compositions (CI through to EL), with the requirement that the embryo has sufficient Fe to match the Earth's present core mass fraction. The dotted lines indicate the initial Fe/Mg for the five chondrite types, the solid black line represents bulk earth if the core contains 10 percent light elements, the dark grey shaded region shows the range for 0--15 percent light element content \citep[as given by][]{ONeill08}. If a range shown for a particular simulation does not overlap with the dotted line indicating the initial Fe/Mg ratio for that chondrite it is not possible to reach the required 32 percent core fraction without stripping of mantle. The upper limits always represent the Fe/Mg achieved from the embryo with maximum increase in core fraction, however the minimum values do not necessarily represent the embryo with the minimum core fraction from that simulation. The lower limit is set by the minimum final core fraction required, for that chondrite type and initial oxidation, to retain sufficient Fe to match the Earth's core fraction, with the requirement that this minimum is within the range of core fractions actually found. If there is no range shown for a simulation the maximum core fraction achieved was not sufficient to reach this minimum required Fe content.
}\label{f:FeMg}}
\end{figure}

{ In order to estimate the bulk Fe/Mg ratios of the final planetary embryos, we use a series of starting bulk compositions for the planetesimals which span the range represented by chondritic meteorites.  Following the technique used in \citetalias{Bonsor15}, for each bulk planetesimal composition, a metallic core is formed of the proscribed mass (either 0.22 or 0.35, see section \ref{s:initialcf}, with 10 percent of the core as ÔlightÕ elements in the former or 5 percent ÔlightÕ elements and 5 percent Si in the latter) and a residual mantle composition calculated.  These compositions are used together with the modelled mass fractions of core and mantle in the final embryos (see Figure \ref{f:all_cf_range}) to determine their bulk Fe/Mg. 
Finally, we apply the additional constraint that the embryo must have enough Fe to match the Earth's measured core mass fraction (the FeO/Fe ratio is adjusted in order to match the Earth's core fraction of 32 percent, ensuring that the FeO content of the mantle is zero or greater).

Figure \ref{f:FeMg} illustrates the differences between starting, chondritic compositions (dashed lines) and bulk Earth composition (dark grey bar) indicating the need for some processing of chondritic material in order to produce the Earth. The ranges in Fe/Mg that result from a selection of our simulations are also shown in Figure \ref{f:FeMg} and it is evident that collisional processing can produce embryos matching Earth's Fe/Mg ratio.  Earth-like Fe/Mg can be produced in extreme cases starting with planetesimal compositions close to bulk Earth in calm scenarios and more readily in the Grand Tack models.  The collisional stripping of mantle caused by the Grand Tack can be sufficient to explain the non-chondritic bulk Fe/Mg ratio of the Earth starting with a wider range of planetesimal compositions.}

It is, however, important to note that the accretion of terrestrial planets is not complete at the end-point of these simulations, and that it is not necessarily the case that the embryos with enhanced core fractions represent the Earth. Collisions during the giant impact phase may cause any compositional differences observed at the end of oligarchic growth to be averaged out again, or they may serve to enhance this variation (see \citealt{Stewart12}; \citetalias{Bonsor15} for more detailed discussions of compositional changes during the giant impact phase).

The compositional variation that we see in these simulations also has strong implications for exoplanetary systems. The bulk composition of planets could be significantly different from the expected composition based on the stellar photosphere. The CI chondrites are a good match for our Sun, but as Figure \ref{f:FeMg} shows the bulk composition of planetary embryos formed from CI chondrites may not be the same. Given the interest in the compositions of exoplanets and their atmospheres, it is important to consider the possible deviation from the composition implied by their host stars, especially if they may have undergone dynamical processes similar to the Grand Tack.
\vspace{5mm}

%%%%%%%%%%%%%%%%%%%% 			SUMMARY 		%%%%%%%%%%%%%%%%%%%%

\section{Summary}

We have investigated the compositional evolution of growing terrestrial planet embryos through runaway and oligarchic growth both under the Grand Tack model \citep{Walsh11} and a model with no influence from giant planets. Our simulations conducted using PKDGRAV \citep{Richardson00,Stadel01} with a state-of-the-art collision model \citep[EDACM,][]{Leinhardt12,Leinhardt15} and mantle stripping laws \citep{Marcus10} show that planetary embryos, grown from differentiated planetesimals with an initially uniform core mass fraction, develop variations in their core fractions, and that both iron-rich and silicate-rich fragments are produced.

In the Grand Tack scenario, and in extreme cases for the calm scenario, these variations in embryo core mass fraction can be sufficient to account for the Earth's non-chondritic Fe/Mg ratio.

In all of the simulations, we see both embryos with increased core fractions and embryos with decreased core fractions. Regardless of whether the compositional shift is in a direction that makes the Earth appear more chondritic, or less chondritic, it is clear that collisions could substantially alter the composition of planetary embryos from that of the initial differentiated planetesimals.

\acknowledgments

We thank the anonymous referee for useful comments that improved this manuscript. PJC would like to thank A.~Bonsor, S.~Lines, J.~Dobinson, S.~Lock and M.~Mace for useful discussions. This work was carried out using the computational facilities of the Advanced Computing Research Centre, University of Bristol - http://www.bris.ac.uk/acrc/. PJC, ZML, TE \& MJW acknowledge support from the Natural Environment Research Council (grant number: NE/K004778/1). ZML also acknowledges support from an STFC Advanced Fellowship.

%\bibliography{../references}

\clearpage

\end{document}